\input harvmac

%
%
%
\message{S-Tables Macro v1.0, ACS, TAMU (RANHELP@VENUS.TAMU.EDU)}
%
%
\newhelp\stablestylehelp{You must choose a style between 0 and 3.}%
\newhelp\stablelinehelp{You should not use special hrules when stretching
a table.}%
\newhelp\stablesmultiplehelp{You have tried to place an S-Table inside another
S-Table.  I would recommend not going on.}%
%
%
\newdimen\stablesthinline
\stablesthinline=0.4pt
\newdimen\stablesthickline
\stablesthickline=1pt
%
%
\newif\ifstablesborderthin
\stablesborderthinfalse
\newif\ifstablesinternalthin
\stablesinternalthintrue
\newif\ifstablesomit
\newif\ifstablemode
\newif\ifstablesright
\stablesrightfalse
%
%
\newdimen\stablesbaselineskip
\newdimen\stableslineskip
\newdimen\stableslineskiplimit
%
%
\newcount\stablesmode
\newcount\stableslines
\newcount\stablestemp
\stablestemp=3
\newcount\stablescount
\stablescount=0
\newcount\stableslinet
\stableslinet=0
%
%
%
\newcount\stablestyle
\stablestyle=0
%
%
\def\stablesleft{\quad\hfil}%
\def\stablesright{\hfil\quad}%
%
%
\catcode`\|=\active%
%
%
\newcount\stablestrutsize
\newbox\stablestrutbox
\setbox\stablestrutbox=\hbox{\vrule height10pt depth5pt width0pt}
\def\stablestrut{\relax\ifmmode%
                         \copy\stablestrutbox%
                       \else%
                         \unhcopy\stablestrutbox%
                       \fi}%
%
%
\newdimen\stablesborderwidth
\newdimen\stablesinternalwidth
\newdimen\stablesdummy
\newcount\stablesdummyc
\newif\ifstablesin
\stablesinfalse
%
%
\def\begintable{\stablestart%
  \stablemodetrue%
  \stablesadj%
  \halign%
  \stablesdef}%
\def\stablesadj{%
  \ifcase\stablestyle%
    \hbox to \hsize\bgroup\hss\vbox\bgroup%
  \or%
    \hbox to \hsize\bgroup\vbox\bgroup%
  \or%
    \hbox to \hsize\bgroup\hss\vbox\bgroup%
  \or%
    \hbox\bgroup\vbox\bgroup%
  \else%
    \errhelp=\stablestylehelp%
    \errmessage{Invalid style selected, using default}%
    \hbox to \hsize\bgroup\hss\vbox\bgroup%
  \fi}%
\def\stablesend{\egroup%
  \ifcase\stablestyle%
    \hss\egroup%
  \or%
    \hss\egroup%
  \or%
    \egroup%
  \or%
    \egroup%
  \else%
    \hss\egroup%
  \fi}%
\def\stablestart{%
  \ifstablesin%
    \errhelp=\stablesmultiplehelp%
    \errmessage{An S-Table cannot be placed within an S-Table!}%
  \fi
  \global\stablesintrue%
  \global\advance\stablescount by 1%
  \message{<S-Tables Generating Table \number\stablescount}%
  \begingroup%
  \stablestrutsize=\ht\stablestrutbox%
  \advance\stablestrutsize by \dp\stablestrutbox%
  \ifstablesborderthin%
    \stablesborderwidth=\stablesthinline%
  \else%
    \stablesborderwidth=\stablesthickline%
  \fi%
  \ifstablesinternalthin%
    \stablesinternalwidth=\stablesthinline%
  \else%
    \stablesinternalwidth=\stablesthickline%
  \fi%
  \tabskip=0pt%
  \stablesbaselineskip=\baselineskip%
  \stableslineskip=\lineskip%
  \stableslineskiplimit=\lineskiplimit%
  \offinterlineskip%
  \def\borderrule{\vrule width \stablesborderwidth}%
  \def\internalrule{\vrule width \stablesinternalwidth}%
  \def\thinline{\noalign{\hrule height \stablesthinline}}%
  \def\thickline{\noalign{\hrule height \stablesthickline}}%
  \def\trule{\omit\leaders\hrule height \stablesthinline\hfill}%
  \def\ttrule{\omit\leaders\hrule height \stablesthickline\hfill}%
  \def\tttrule##1{\omit\leaders\hrule height ##1\hfill}%
  \def\stablesel{&\omit\global\stablesmode=0%
    \global\advance\stableslines by 1\borderrule\hfil\cr}%
  \def\el{\stablesel&}%
  \def\elt{\stablesel\thinline&}%
  \def\eltt{\stablesel\thickline&}%
  \def\elttt##1{\stablesel\noalign{\hrule height ##1}&}%
  \def\elspec{&\omit\hfil\borderrule\cr\omit\borderrule&%
              \ifstablemode%
              \else%
                \errhelp=\stablelinehelp%
                \errmessage{Special ruling will not display properly}%
              \fi}%
  \def\stmultispan##1{\mscount=##1 \loop\ifnum\mscount>3 \stspan\repeat}%
  \def\stspan{\span\omit \advance\mscount by -1}%
  \def\multicolumn##1{\omit\multiply\stablestemp by ##1%
     \stmultispan{\stablestemp}%
     \advance\stablesmode by ##1%
     \advance\stablesmode by -1%
     \stablestemp=3}%
  \def\multirow##1{\stablesdummyc=##1\parindent=0pt\setbox0\hbox\bgroup%
    \aftergroup\emultirow\let\temp=}
  \def\emultirow{\setbox1\vbox to\stablesdummyc\stablestrutsize%
    {\hsize\wd0\vfil\box0\vfil}%
    \ht1=\ht\stablestrutbox%
    \dp1=\dp\stablestrutbox%
    \box1}%
  \def\stpar##1{\vtop\bgroup\hsize ##1%
     \baselineskip=\stablesbaselineskip%
     \lineskip=\stableslineskip%
     \lineskiplimit=\stableslineskiplimit\bgroup\aftergroup\estpar\let\temp=}%
  \def\estpar{\vskip 6pt\egroup}%
  \def\stparrow##1##2{\stablesdummy=##2%
     \setbox0=\vtop to ##1\stablestrutsize\bgroup%
     \hsize\stablesdummy%
     \baselineskip=\stablesbaselineskip%
     \lineskip=\stableslineskip%
     \lineskiplimit=\stableslineskiplimit%
     \bgroup\vfil\aftergroup\estparrow%
     \let\temp=}%
  \def\estparrow{\vfil\egroup%
     \ht0=\ht\stablestrutbox%
     \dp0=\dp\stablestrutbox%
     \wd0=\stablesdummy%
     \box0}%
  \def|{\global\advance\stablesmode by 1&&&}%
  \def\|{\global\advance\stablesmode by 1&\omit\vrule width 0pt%
         \hfil&&}%
  \def\vt{\global\advance\stablesmode by 1&\omit\vrule width \stablesthinline%
          \hfil&&}%
  \def\vtt{\global\advance\stablesmode by 1&\omit\vrule width
\stablesthickline%
          \hfil&&}%
  \def\vttt##1{\global\advance\stablesmode by 1&\omit\vrule width ##1%
          \hfil&&}%
  \def\vtr{\global\advance\stablesmode by 1&\omit\hfil\vrule width%
           \stablesthinline&&}%
  \def\vttr{\global\advance\stablesmode by 1&\omit\hfil\vrule width%
            \stablesthickline&&}%
  \def\vtttr##1{\global\advance\stablesmode by 1&\omit\hfil\vrule width
##1&&}%
  \stableslines=0%
  \stablesomitfalse}
\def\stablesdef{\bgroup\stablestrut\borderrule##\tabskip=0pt plus 1fil%
  &\stablesleft##\stablesright%
  &##\ifstablesright\hfill\fi\internalrule\ifstablesright\else\hfill\fi%
  \tabskip 0pt&&##\hfil\tabskip=0pt plus 1fil%
  &\stablesleft##\stablesright%
  &##\ifstablesright\hfill\fi\internalrule\ifstablesright\else\hfill\fi%
  \tabskip=0pt\cr%
  \ifstablesborderthin%
    \thinline%
  \else%
    \thickline%
  \fi&%
}%
\def\endtable{\advance\stableslines by 1\advance\stablesmode by 1%
   \message{- Rows: \number\stableslines, Columns:  \number\stablesmode>}%
   \stablesel%
   \ifstablesborderthin%
     \thinline%
   \else%
     \thickline%
   \fi%
   \egroup\stablesend%
\endgroup%
\global\stablesinfalse}
%


\def\rappendix#1#2{\global\meqno=1\global\subsecno=0\xdef\secsym{\hbox{#1.}}
\bigbreak\bigskip\noindent{\bf Appendix. #2}\message{(#1. #2)}
\writetoca{Appendix {#1.} {#2}}\par\nobreak\medskip\nobreak}

\def\ack{\bigbreak\bigskip\centerline{{\bf Acknowledgements}}\nobreak
\bigskip}

\sfcode`\?=3000

%
%


\input epsf \def\ynepsf#1#2{\epsfxsize=#2\epsfbox{#1}}



%

\newcount\figno \figno=0

\def\insertfig#1#2#3{\midinsert\centerline{\ynepsf{#1}{#2}}
\global\advance\figno by1
\setbox0=\vbox{\noindent\leftskip=.2\hsize\rightskip=.2\hsize
{\bf Figure \the\figno}. #3}
\vskip 18pt
\centerline{\box0\hss}
\vskip 24pt
\endinsert}


\def\insertfigtex#1#2{
\midinsert\centerline{\vbox{
\begintable
#1
\endtable
\global\advance\figno by1
\setbox0=\vbox{\noindent\leftskip=.2\hsize\rightskip=.2\hsize
{\bf Figure \the\figno}. #2}
\vskip 18pt
\centerline{\box0\hss}
}}\endinsert}

\newcount\tabno \tabno=0

\def\inserttabtex#1#2{
\midinsert\centerline{\vbox{
\begintable
#1
\endtable
\global\advance\tabno by1
\setbox0=\vbox{\noindent\leftskip=.2\hsize\rightskip=.2\hsize
{\bf Table \the\tabno}. #2}
\vskip 18pt
\centerline{\box0\hss}
\vskip 24pt
}}\endinsert}


\def\MPL#1#2#3{{Mod.~Phys.~Lett.} {\bf A#1} (19#2) #3}

\def\PL#1#2#3{{Phys.~Lett.} {\bf B#1} (19#2) #3}
\def\NP#1#2#3{{Nucl.~Phys.} {\bf B#1} (19#2) #3}
\def\CMP#1#2#3{{Commun.~Math.~Phys.} {\bf #1} (19#2) #3}
\def\LMP#1#2#3{{Lett.~Math.~Phys.} {\bf #1} (19#2) #3}
\def\TMP#1#2#3{{Theor.~Math.~Phys.} {\bf #1} (19#2) #3}
\def\TMF#1#2#3{{Teor.~Mat.~Fiz.} {\bf #1} (19#2) #3}
\def\JETP#1#2#3{{Sov.~Phys.~JETP.} {\bf #1} (19#2) #3}
\def\ZETF#1#2#3{{Zh.~Eksp.~Teor.~Fiz.} {\bf #1} (19#2) #3}


\def\frac#1#2{{#1 \over #2}}
\def\inv#1{{1 \over #1}}
\def\ha{{1 \over 2}}
\def\der{\partial}
\def\vev#1{\left\langle #1 \right\rangle}
\def\dvev#1{\left\langle\!\left\langle #1 \right\rangle\!\right\rangle}
\def\ddvev#1{\left\langle\!\!\left\langle #1 \right\rangle\!\!\right\rangle}
\def\dddvev#1{\left\langle\!\!\!\left\langle #1 \right\rangle\!\!\!\right
\rangle}
\def\ket#1{ \vert#1\rangle}
\def\->{\rightarrow}     \def\<-{\leftarrow}
\def\<{\langle}          \def\>{\rangle}
\def\[{\left [}          \def\]{\right ]}
\def\({\left (}          \def\){\right )}
\def\makeblankbox#1#2{\hbox{\lower\dp0\vbox{\hidehrule{#1}{#2}%
   \kern -#1
   \hbox to \wd0{\hidevrule{#1}{#2}%
      \raise\ht0\vbox to #1{}
      \lower\dp0\vtop to #1{}
      \hfil\hidevrule{#2}{#1}}%
   \kern-#1\hidehrule{#2}{#1}}}%
}%
\def\hidehrule#1#2{\kern-#1\hrule height#1 depth#2 \kern-#2}%
\def\hidevrule#1#2{\kern-#1{\dimen0=#1\advance\dimen0 by #2\vrule
    width\dimen0}\kern-#2}%
\def\openbox{\ht0=1.2mm \dp0=1.2mm \wd0=2.4mm  \raise 2.75pt
\makeblankbox {.25pt} {.25pt}  }
\def\qed{\hskip 8mm \openbox}
\def\abs#1{\left\vert #1 \right\vert}
\def\bun#1/#2{\leavevmode
   \kern.1em \raise .5ex \hbox{\the\scriptfont0 #1}%
   \kern-.1em $/$%
   \kern-.15em \lower .25ex \hbox{\the\scriptfont0 #2}%
}
\def\row#1#2{#1_1,\ldots,#1_#2}
\def\apar{\noalign{\vskip 2mm}}
\def\blackbox{\hbox{\vrule height .5ex width .3ex depth -.3ex}}
\def\nord{{\textstyle {\blackbox\atop\blackbox}}}
\def\ts{\,}


\def\op#1{\mathop{\rm #1}\nolimits}
\def\dim{\op{dim}}
\def\min{\op{min}}


\def\bg{{\bf g}}

\def\bZ{{\bf Z}}


\def\a{\alpha}
\def\b{\beta}
\def\c{\chi}
\def\d{\delta}		\def\D{\Delta}
\def\e{\epsilon}
\def\g{\gamma}		\def\G{\Gamma}
\def\la{\lambda}	        \def\La{\Lambda}
\def\m{\mu}
\def\n{\nu}
\def\o{\omega}		
\def\p{\phi}            \def\P{\Phi}
\def\ps{\psi}           \def\Ps{\Psi}
\def\r{\rho}
	
\def\t{\theta}		
\def\vp{\varphi}
\def\z{\zeta}

%
\def\CA{{\cal A}}	\def\CB{{\cal B}}	\def\CC{{\cal C}}
\def\CD{{\cal D}}		\def\CF{{\cal F}}
\def\CG{{\cal G}}		
\def\CJ{{\cal J}}		
	\def\CN{{\cal N}}	
\def\CP{{\cal P}}		\def\CR{{\cal R}}
	\def\CT{{\cal T}}	
\def\CV{{\cal V}}	\def\CW{{\cal W}}	
	
%


\def\minimal{$N=2$ minimal model}
\def\eb{{\bar\eta}}

\hyphenation{functions}
\hyphenation{para-fermions}

\lref\rA{K.~Aomoto, in {\it Ramanujuan Revisited},
          Proceedings of the Centenary Conference, Eds.~G.F.~Andrews et al.,
         (Academic Press, New York, 1988).}
\lref\rB{N.~Bourbaki, {\it Groupes et alg{\`e}bres de Lie}, (Masson,
Paris, 1981).}
\lref\rBMP{P.~Bouwknegt, J.~McCarthy and K.~Pilch, \CMP{131}{90}{125}.}
\lref\rD{E.~Date, M.~Jimbo, A.~Kuniba, T.~Miwa and M.~Okado,
\LMP{17}{89}{69}.}
\lref\rDF{Vl.S.~Dotsenko and V.A.~Fateev, \NP{251}{85}{691}.}
\lref\rDHS{G.~Dunne, I.~Halliday and P.~Suranyi, \NP{325}{89}{526}.}
\lref\rDVV{R.~Dijkgraaf, E.~Verlinde and H.~Verlinde, \NP{352}{91}{59};
           PUPT-1217, IASSNS-HEP-90-80, (November, 1990).}
\lref\rEHY{T.~Eguchi, S.~Hosono and S.-K.~Yang, \CMP{140}{91}{159}.}
\lref\rEKMY{T.~Eguchi, T.~Kawai, S.~Mizoguchi and S.-K.~Yang,
            KEK-TH-303, KEK preprint 91-111, (September 1991), to appear in
            Rev.~Math.~Phys.}
\lref\rEY{T.~Eguchi and S.-K.~Yang, \MPL{4}{90}{1653}. }
\lref\rF{G.~Felder, \NP{317}{89}{215}.}
\lref\rFZ{V.A.~Fateev and A.B.~Zamolodchikov, \NP{{280 [FS18]}}{87}{644}.}%
\lref\rG{D.~Gepner, \NP{290}{87}{10}.}
\lref\rGii{D.~Gepner, \CMP{141}{91}{381}.}
\lref\rGoverG{%
K.~Gawedzki and A.~Kupiainen, \NP{320}{89}{625}\semi
H.~Verlinde, \NP{337}{90}{652}\semi
E.~Witten, \CMP{144}{92}{189}\semi
K.~Intriligator, \MPL{6}{91}{3543}\semi
M.~Spiegelglas and S.~Yankielowicz, TECHNION-PH-90-34.}
\lref\rGNOS{P.~Goddard, N.~Nahm, D.~Olive and A.~Schwimmer,
\CMP{107}{86}{179}.}
\lref\rGS{P.~Goddard and A.~Schwimmer, \PL{206}{88}{62}.}
\lref\rI{K.~Intriligator, \MPL{6}{91}{3543}.}
\lref\rK{T.~Kawai, \PL{259}{91}{460}; {\bf B261} (1991) 520E.}
\lref\rKP{V.G.~Kac and D.H.~Peterson, Adv.~Math. {\bf 53} (1984) 125.}
\lref\rKUY{T.~Kawai, T.~Uchino and S.-K.~Yang,
            KEK-TH-311, KEK preprint 91-190, (December 1991), to appear in
             Prog.~Theor.~Phys.~Suppl.}
\lref\rL{W.~Lerche, \PL{252}{90}{349}.}
\lref\rLVW{W.~Lerche, C.~Vafa and N.P.~Warner, \NP{324}{89}{427}.}
\lref\rMinimal{A.B.~Zamolodchikov and V.A.~Fateev, \JETP{63}{86}{913}
             \hfil\break
            [\ZETF{90}{86}{1553}]\semi
W.~Boucher, D.~Friedan and A.~Kent, \PL{172}{86}{316}\semi
P.~Di Vecchia, J.L.~Petersen, M.~Yu and H.B.~Zheng, \PL{174}{86}{280}\semi
Z.~Qiu, \PL{188}{87}{207}.}
\lref\rNW{D.~Nemeschansky and N.P.~Warner, USC-91/031, (October, 1991).}
\lref\rNY{M.~Ninomiya and K.~Yamagishi, \PL{183}{87}{323}.}
\lref\rOthers{P.~Bouwknegt, J.~McCarthy and K.~Pilch,
       \CMP{131}{90}{125}\semi
S.~Mizoguchi and T.~Nakatsu, Tokyo preprint, UT566, (September 1990)\semi
S.-w.~Chung, E.~Lyman and S.-H.~H.~Tye,
            Cornell preprint, CLNS 91/1057, (May 1991).}
\lref\rR{L.J.~Romans, \NP{352}{91}{829}.}
\lref\rS{H.~Saleur,  YCTP-p38-91 and YCTP-p39-91 (October, 1991).}
\lref\rSS{A.~Schwimmer and N.~Seiberg, \PL{184}{87}{191}.}
\lref\rW{E.~Witten, \CMP{118}{88}{411}; \NP{340}{90}{281}.}
\lref\rZFi{A.B.~Zamolodchikov and V.A.~Fateev, \JETP{62}{85}{215}
             \hfil\break
            [\ZETF{89}{85}{380}].}
\lref\rZFii{A.B.~Zamolodchikov and V.A.~Fateev, \JETP{63}{86}{913}
             \hfil\break
            [\ZETF{90}{86}{1553}].}
\lref\rZFiii{A.B.~Zamolodchikov and V.A.~Fateev, \TMP{71}{88}{451}
           \hfil\break
          [\TMF{71}{87}{163}].}


\noblackbox
\vbadness=10000
\Title{\vbox{\baselineskip12pt\hbox{KEK-TH-336}
                             \hbox{KEK preprint 92-42}
                             \hbox{hep-th/9206110}
                              \hbox{June 1992}}}
{
\vbox{\vskip -3cm\centerline{Higher-Rank Supersymmetric Models and}
\medskip
\centerline{Topological  Conformal Field Theory}}}

\centerline{%
Toshiya Kawai${}^\ast$, Taku Uchino${}^\dagger$ and Sung-Kil Yang${}^\ddagger$
\footnote{${}^\$ $}{Partially supported by Grant-in-Aid for Scientific
Research
on Priority Area 231 ``Infinite Analysis''.}
}%
\it
\vskip.2in
\centerline{${}^\ast$National Laboratory for High Energy Physics (KEK)}
\centerline{Tsukuba, Ibaraki 305, Japan}
\medskip
\centerline{${}^\dagger$Department of Physics, Tokyo Institute of Technology,}
\centerline{Oh-okayama, Meguro-ku, Tokyo 152, Japan}
\medskip
\centerline{${}^\ddagger$Institute of Physics, University of Tsukuba,}
\centerline{Tsukuba, Ibaraki 305, Japan}
\rm
\vskip .2in
\centerline{\bf Abstract}
\medskip
In the first part of this paper we investigate the operator aspect of
higher-rank supersymmetric model which is  introduced as a Lie theoretic
extension of the \minimal\ with the simplest case $su(2)$ corresponding to
the \minimal.
 In particular we identify the analogs of chirality conditions and chiral
ring.
In the second part we construct a class of topological conformal field
theories
starting with this higher-rank supersymmetric model.
We show the BRST-exactness of the twisted stress-energy tensor,
find out physical observables  and discuss how to
make their  correlation functions.
It is emphasized that in the case of $su(2)$ the topological field theory
constructed in this paper is distinct from the one obtained by twisting
the \minimal\  through the usual procedure.

\Date{}
\footline={\hss\tenrm-- \folio\ --\hss}

\newsec{Introduction}

Over the past several  years the \minimal\ \rMinimal\
has been  a useful testing grounds for various theoretical ideas, especially
those of geometrical and topological origins.
It is natural to ask whether one can invent  any interesting generalizations
of the \minimal\ and, if such generalizations ever exist,
what kind of insights  one can gain from them.
As discussed in  \refs{\rK,\rEKMY}, it is not difficult to imagine
 a class of models described by a set of free bosons and parafermions
associated with an arbitrary simple Lie algebra $\bg$ which reduce to the
$N=2$ minimal model when specialized to $\bg=su(2)$.
The approach taken there was  primarily through branching relations.
A simple analysis on these relations suggested the existence of
`supersymmetry'  generalizing $N=2$ superconformal symmetry and the spectra
of (analogs of) chiral primary states and the Ramond ground states were
derived.
We will refer to these models as higher-rank supersymmetric models.

The first aim of this article is to elaborate on the operator aspect
of these models.  We first review some basic properties of
the WZW model and the associated parafermions in
sect.2. In sect.3, after recalling the construction of the models, we consider
the structure of chiral algebra and spectral flow.
Then  we  investigate the chiral primary state conditions and the chiral ring
together with their connections to the Ramond ground states.
These results are mostly direct generalizations
of the essential  features of the \minimal, however, we will find that
some properties observed in the \minimal\ are  due to the specific nature
of $su(2)$.

One of the important characteristics  of the \minimal\ is that they can
be `twisted' yielding  a class of topological field theories \rW\ as
discussed in \rEY. This line of thought was further pursued in
\refs{\rL,\rEHY}.
Furthermore, it was investigated in \rEKMY\ for $N=2$
hermitian symmetric space
models how the twisting procedure can be seen at the level of characters.
A natural question would be whether we can similarly make
topological field theory out of our higher-rank supersymmetric model.
Investigating this problem is the  second aim of the paper.
As it turns out in sect.4 the answer to this question is a qualified yes:
topological field theory we are able to construct for an arbitrary
$\bg$ is {\it not\/} strictly  of the type we usually associate with the
\minimal.
Our reasoning  rests critically on the BRST-exact expression of
the twisted stress-energy tensor originally proposed in \rEHY\ in the context
of $c=0$ GKO coset models. Actually a large portion of this paper  is
concerned
with the (partial) proof of this formula.
The next step is to seek  BRST invariant
physical observables and study their correlation functions.
To produce viable correlation functions among the basic physical observables
we have to introduce extra BRST invariant operators.
As a consequence of this  the topological conformal
field theory we consider here for $\bg=su(2)$ is distinct from  the one for
the \minimal\ in the ordinary sense.
We then comment on perturbations of our topological field theory and finally
 will  speculate on the relation to $G/G$ theory \rGoverG.

In sect.5 concluding remarks are given while
appendix contains several formulas to be used in sect.4 in evaluating
multiple integrals.

A preliminary account of this work was reported in \rKUY.

\newsec{WZW model and Parafermions.}

In this section we briefly summarize some basic
properties of the WZW model and the associated parafermions
\refs{\rZFi\rNY{--}\rG}\
for the sake of later convenience.

We start with fixing our notation for  Lie algebras.
Let $\bg$ be a finite-dimensional simple Lie algebra of rank $n$ and let
 $\row \a n$ be its simple roots. The set of positive roots of $\bg$ is
denoted
by $\D_+$ so that $\D=\D_+\cup(-\D_+)$ is the whole set of roots.
We always follow the convention such that any long root of $\bg$  has length
$\sqrt{2}$ and hence $\t^2=2$ where $\t$ is the highest root of $\bg$.
For any root $\a$ its coroot is defined by $\a^\vee=\frac{2}{\a^2}\ts\a$.
Define a symmetric matrix $G$ by $(G)_{ij}=\a_i^\vee\cdot\a_j^\vee$.
The fundamental weights $\row\o n$ are dual to the simple coroots:
$\o_i\cdot\a_j^\vee=\d_{ij}$.
Let $P$ be the weight lattice of $\bg$ and let $M$ be  the coroot lattice of
$\bg$. The Weyl vector $\r$ is defined to be
half the sum of the positive roots.

The WZW model at level $k$, where $k$ is a positive integer, is algebraically
described by the integrable representations at level $k$ of the (untwisted)
affine Kac-Moody algebra associated with $\bg$. In the Chevalley basis
 the Kac-Moody algebra is  determined by the mutual operator product
expansions of $H_i(z),\ i=1,\ldots,n$ and $E_\a(z),\ \a\in\D$:
\eqna\kmOPE
$$\eqalignno{
&H_i(z)H_j(z')=\frac{k\ts(G)_{ij}}{(z-z')^2}+\cdots\ts ,     &\kmOPE a\cr
\apar
&H_i(z)E_\a(z')=\frac{(\a\cdot\a_i^\vee)}{z-z'}E_\a(z')+\cdots\ts, &\kmOPE
b\cr
\apar
&E_\a(z)E_{-\a}(z')=\frac{k\ts(2/\a^2)}{(z-z')^2}+
              \inv{z-z'}\sum_{i=1}^n \n_i H_i(z')+\cdots\ts,  &\kmOPE c\cr
\apar
&E_\a(z)E_\b(z')=\frac{N_{\a,\b}}{z-z'}E_{\a+\b}(z')+\cdots\ts.
                                                              &\kmOPE d\cr}$$
In eq.\kmOPE{c}\ $\n_i$'s are the integers such that
$\a^\vee=\sum_{i=1}^n \n_i \a_i^\vee$ and in eq.\kmOPE{d}\ $N_{\a,\b}=0$ if
$\a+\b \not\in\D$
and $N_{\a,\b}=\pm(s+1)$ if $\a+\b\in \D$ where $s$ is the largest integer
such that $\a-s\b\in \D$.

The stress-energy tensor of the WZW model in the Sugawara form is
\eqnn\sug
$$\eqalignno{&T_{\rm Sug}(z)=\inv{2(k+g)}\[\sum_{i,j=1}^n(G^{-1})_{ij}
\nord H_i(z)H_j(z)\nord+\sum_{\a\in\D}\frac{\a^2}{2}\nord E_\a(z)E_{-\a}(z)
\nord\]\,,\cr
\apar
&&\sug\cr}
$$
where $g$ is the dual Coxeter number of $\bg$. As is well-known, the moments
of $T_{\rm Sug}(z)$ generate the Virasoro algebra with a  central charge
\eqn\ccharge{c=\frac{kd}{k+g}\ts ,}
where $d=\dim\bg$.

The generators of the Kac-Moody algebra can be  realized
in terms of free bosons $\vp(z)=(\vp_1(z),\ldots,\vp_n(z))$
and parafermions $\ps_\a(z),\ \a\in\D$  \refs{\rNY,\rG}:
\eqn\KMgen{\eqalign{
&H_i(z)=i\sqrt{k}\a_i^\vee\cdot\der\vp(z), \quad i=1,\ldots,n\ts , \cr
\apar
&E_\a(z)=\sqrt{\frac{2k}{\a^2}}
            \ps_\a(z)  \nord\exp\[i\a\cdot\vp(z)/\sqrt{k}\]\!\nord\,
            c_\a\ts,\quad\a\in\D\ts ,\cr}}
where  the cocycle factors $c_\a$ commute with $\vp(z)$
and $\ps_\a(z)$,  satisfying
\eqn\?{c_\a c_\b=\e(\a,\b)c_{\a+\b}\ts,\quad\e(\a,\b)=\pm 1\ts,}
with
\eqn\?{\eqalign{&\e(\a,\b)\e(\a+\b,\g)=\e(\a,\b+\g)\e(\b,\g)\ts,\cr
                       &\e(\a,\b)=(-1)^{\a\cdot\b+\a^2\b^2}\e(\b,\a)\ts.\cr}}
The OPEs of the free bosons and the parafermions are
\eqna\bpfOPE{
$$\eqalignno{
&\vp_a(z)\vp_b(z')=-\d_{ab}\log(z-z')+\cdots\ts ,&\bpfOPE a\cr
\apar
&\ps_\a(z)\ps_{-\a}(z')=\inv{(z-z')^{2-\a^2/k}}\[I+
                     (z-z')^2\ts T_\a^{\rm (pf)}(z')+\cdots\]\ts,&\bpfOPE b\cr
\apar
&\ps_\a(z)\ps_\b(z')=\frac{K(\a,\b)}{(z-z')^{1+\a\cdot\b/k}}
      \ps_{\a+\b}(z')+\cdots\ts, &\bpfOPE c\cr}$$}
where
\eqn\pfcoeff{
K(\a,\b)\e(\a,\b)=N_{\a,\b}\ts\sqrt{ \frac{\a^2 \b^2}{2k(\a+\b)^2} }\ts.}
One can assume without loss of generality $\e(\a,0)=\e(0,\a)=\e(\a,-\a)=1$
so that $\e(\a,\b)=\e(-\b,-\a)$.

Substituting \KMgen\ in \sug, we find that
\eqna\stress
$$\eqalignno{
&T_{\rm Sug}(z)=-\ha\nord\der\vp(z)\cdot\der\vp(z)\nord+T^{\rm (pf)}(z)\ts,
&\stress a\cr
\apar
&T^{\rm (pf)}(z):=\frac{k}{k+g}\sum_{\a\in\D_+}\ts T_\a^{\rm (pf)}(z)
\ts, &\stress b\cr}$$
where we used
\eqn\aortho{\sum_{\a\in\D_+}(\a)_a(\a)_b=g\d_{ab}\ts,}
and $T_\a^{\rm (pf)}(z)=T_{-\a}^{\rm (pf)}(z)$.

We now wish to recall the interacting boson representation of parafermions
\rDHS\ as a  preparation for later use. For simplicity, assume $\bg$ to be
simply-laced\foot{
The Dynkin diagram of any non simply-laced Lie algebra can be obtained by
folding that  of an appropriate simply-laced Lie algebra.
Using this fact, one can find \rGNOS\
a vertex operator construction for the level 1 representation of the affine
Lie algebra associated with an arbitrary non simply-laced Lie algebra.
Hence there is no serious obstacle in extending the results
in \rDHS\ for
a non simply-laced $\bg$.}
and let $\c_a^b(z)$ $(a=1,\ldots, n,\ b=1,\ldots,k-1)$ be free bosons
normalized
 by $\vev{\c_a^b(z)\c_{a'}^{b'}(z')}=-\d_{aa'}\d_{bb'}\log(z-z')$.
Let $\row\la k$ be the  weights in the vector representation of $su(k)$ which
can be considered as $(k-1)$-dimensional vectors satisfying
\eqn\sunvec{
\sum_{j=1}^k\la_j=0,\quad \la_i\cdot\la_j=\d_{ij}-\inv{k},\quad
\sum_{i=1}^k(\la_i)_a(\la_i)_b=\d_{ab}\ts.}
Notice that $\D(su(k))=\{\la_i-\la_j\ \vert\ i\ne j\}$ where $\D(su(k))$
is the set of roots of $su(k)$.
Then the parafermions can be represented by \rDHS
\eqn\pfbyboson{\ps_\a(z)c_\a=
      \inv{\sqrt{k}}\sum_{j=1}^k \nord\exp[i\a \otimes\la_j\cdot\c(z)]\nord
c_\a^j\ts,}
where $c_\a^j$'s are  $k$ copies of the cocycle
factors of $\bg$.
In this representation,
\eqnn\?
$$\eqalignno{&T_\a^{\rm (pf)}(z)=\inv{k}\[-\ha \nord(\a\cdot\der\c(z))\cdot
(\a\cdot\der\c(z))\nord+\sum_{\b\in\D(su(k))}
\nord\exp[i\a\otimes\b\cdot\c(z)]\nord\,c^\b_\a\]\,,\cr
\apar
&&\?\cr}$$
and hence
\eqnn\pfstboson
$$\eqalignno{&T^{\rm (pf)}(z)=\inv{k+g}\[-\frac{g}{2}\,
\nord\der\c(z)\cdot\der\c(z)\nord
+\sum_{\a\in\D_+}\sum_{\b\in\D(su(k))}
\nord\exp[i\a\otimes\b\cdot\c(z)]\nord\,c^\b_\a \]\,,\cr
\apar
&&\pfstboson\cr}$$
where $c_\a^\b=c_\a^ic_{-\a}^j$ for $\b=\la_i-\la_j$.

Let $P_+^k =\{\La \in \sum_{i=1}^n \bZ_{\geq 0}\,\omega_i
:\La \cdot \theta \leq k \}$. The primary fields of the WZW model are given by
\eqn\?{(\P_{\rm WZW})^\La_\la(z)=\p^\La_\la(z)
\nord\exp\[i\la\cdot\vp(z)/\sqrt{k}\]\nord\,,\qquad \La\in P_+^k\,,}
where $\la$ is a weight in the highest weight representation of $\bg$
with highest weight $\La$ and $\p^\La_\la(z)$ is a parafermionic field
of conformal weight
\eqn\?{\frac{\La\cdot(\La+2\r)}{2(k+g)}-\frac{\la^2}{2k}\,.}
The field $\p^\La_\la(z)$  can be obtained from the parafermionic primary
field
$\p^\La_\La(z)$ by repeatedly taking OPEs with
$\ps_{-\a}(z)$, $\a\in\D_+$. The  conformal  weight of
$(\P_{\rm WZW})^\La_\la(z)$ is $h_\La= \frac{\La\cdot(\La+2\r)}{2(k+g)}$.

We now comment on the notion of `charge conjugation' in the WZW model.
It is convenient to put the theory on a torus.
The $\CP\CT$ transformation on a torus is given by
$S^2$ where $S$ is the modular transformation $\tau\->-1/\tau$ (since
$S^2$ reverses the orientations of the  two homologically non-trivial cycles
on the torus.)
Consequently, by $\CC\CP\CT$ invariance, $S^2=\CC$ where $\CC$ is the charge
conjugation
transformation  satisfying $\CC^2=1$.
Let $W$ be the Weyl group of $\bg$ and  let $\ell(w)$ denote
the length of $w$ for each $w\in W$. There exists a unique longest element
$w_0$
in $W$ satisfying \rB
\eqn\longweyl{
\eqalign{
&\ell(w_0)=\abs{\D_+}\,,\quad w_0^2={\rm id}\,,\cr
&\ell(ww_0)=\ell(w_0w)=\ell(w_0)-\ell(w) \quad\hbox{for all $w\in W$}\,,\cr
&w_0(\a_i)=-\a_{p(i)}\,, \quad (i=1,\ldots,n)\quad
           \hbox{for some permutation $p$}\,,\cr
&w_0(\t)=-\t\,,\qquad w_0(\r)=-\r\,.\cr}}
The conformal families of the WZW model are labeled by $P_+^k$ and
 the conformal blocks on a torus are  the Weyl-Kac characters
$\c_\La(\tau)$, $\La\in P_+^k$.
The modular transformation $S$ is given by \rKP
\eqn\?{(S)_{\La,\La'}=
(i)^{\abs{\D_+}}\abs{P/(k+g)M}^{-1/2}\sum_{w\in W}(-1)^{\ell(w)}
e^{-\frac{2\pi i}{k+g} w(\La+\r)\cdot(\La'+\r)}\,.}
Noting that $S$ is a symmetric and unitary matrix, it follows that
$S^*=S\CC=\CC S$. Using this and \longweyl\ it is easy to see that
\eqn\?{(\CC)_{\La,\La'}=\d_{\bar\La,\La'}\,,}
where $\bar\La=-w_0(\La)$ is the highest weight in the complex conjugate
representation of the irreducible representation of $\bg$ with highest weight
$\La$. Note that $\La\in P_+^k \Leftrightarrow \bar\La\in P_+^k$ and
$h_{\bar\La}=h_\La$ by virtue of \longweyl.
Therefore the field conjugate to $(\P_{\rm WZW})^\La_\la(z)$
is given by $(\P_{\rm WZW})^{\bar\La}_{-\la}(z)$ satisfying
\eqn\?{\vev{(\P_{\rm WZW})^\La_\la(z)(\P_{\rm WZW})^{\bar\La}_{-\la}(z')}=
\inv{(z-z')^{2h_\La}}\,.}

Before closing this section, we should mention some auxiliary facts to be used
in the next section.
Notice that
\eqn\acomp{\nord E_{-\a}(z)E_\a(z)\nord=T_\a^{\rm (pf)}(z)-
\inv{2k}\nord(\a\cdot\der\vp(z))^2\nord-
\inv{2\sqrt{k}}i\a\cdot\der^2\vp(z)\,.}
Define $[X]_m$ for any field $X(z)$ of conformal weight
$\D_X$  and for each $m\in \bZ$ by $X(z)=\sum_{m\in\bZ}[X]_m z^{-m-\D_X}$.
{}From  $[E_\a]_m(\P_{\rm WZW})^\La_\La(0)\vert 0\rangle=0$, $m\ge 0$,
$\a\in \D_+$, we observe that
\eqn\acompann{[\nord E_{-\a}E_\a\nord]_0
(\P_{\rm WZW})^\La_\La(0)\vert 0\rangle=0\,,\qquad \a\in\D_+\,.}
Then it follows from \acomp\ and \acompann\ that
\eqn\eigenpf{[T_\a^{\rm (pf)}]_0\p^\La_\La(0)\vert 0\rangle=
\(\inv{2k}\a\cdot\La-\inv{2k^2}(\a\cdot\La)^2\)\p^\La_\La(0)\vert 0\rangle\,,
\qquad\a\in\D_+\,.}
As a consistency check we find
\eqnn\?
$$\eqalignno{
[T^{\rm (pf)}]_0\p^\La_\La(0)\vert 0\rangle&=
\frac{k}{k+g}\sum_{\a\in\D_+}\(\inv{2k}\a\cdot\La-\inv{2k^2}(\a\cdot\La)^2\)
\p^\La_\La(0)\vert 0\rangle\cr
\apar
&=\(\frac{\La\cdot(\La+2\r)}{2(k+g)}-\frac{\La^2}{2k}\)
\p^\La_\La(0)\vert 0\rangle\,.&\?\cr}
$$

\newsec{Higher-Rank Supersymmetric Models}

We first recall the basic features of the higher-rank supersymmetric models
\refs{\rK,\rEKMY}.
The model consists of free bosons and parafermions in much the same way as in
the WZW model. The difference arises from a particular choice of the Coulomb
gas parameters. We set them as follows:
\eqn\?{\a_+=\sqrt{\frac{k+g}{kg}}\,,\quad\a_+\a_-=-\inv{k}\,,\quad
\a_0=\a_++\a_-\,.}
The stress-energy tensor of the model is
\eqn\?{\eqalign{T(z)&=-\ha\nord(\der\vp(z))^2\nord+T^{\rm (pf)}(z)\cr
\apar
&=\sum_{\a\in\D_+}T_\a(z)=\sum_{m\in\bZ}z^{-m-2}L_m\,,\cr}}
where
\eqn\Trep{T_\a(z)=\frac{c}{d}\[\ha\nord(i\a_+\a\cdot\der\vp(z))^2\nord+
T_\a^{\rm(pf)}(z)\]=\sum_{m\in\bZ}z^{-m-2}L_m^\a\,,}
and  $T(z)$ satisfies the Virasoro algebra with the same central charge as
the WZW model, namely, the value $c$ given by \ccharge.
The fundamental fields in the model are
\eqn\?{\Ps^{\La,\m}_\n(z)=\p^\La_{\n-\m}(z)
\nord\exp[-i(\a_+\m+\a_-\n)\cdot\vp(z)]\nord\,,\quad \La\in P_+^k\,,}
where  $\m\in P/gM$ and $\n\in P/(k+g)M$ are such that $\n-\m$ is a weight in
the irreducible representation of $\bg$ with highest weight $\La$.
The conformal weight of $\Ps^{\La,\m}_\n(z)$ is
\eqn\?{\eqalign{&\frac{\La\cdot(\La+2\r)}{2(k+g)}-\frac{(\n-\m)^2}{2k}+
\ha(\a_+\m+\a_-\n)^2\cr
&\hskip 2.5cm =\frac{\La\cdot(\La+2\r)-\n^2}{2(k+g)}+\frac{\m^2}{2g}\,.\cr}}

The  higher-rank supercurrents are defined by
\eqn\Grep{G^\a(z)=\sqrt{\frac{c}{d}}\ps_\a(z)\nord
e^{i\a_+\a\cdot\vp(z)}\nord\,,
\quad\a\in\D\,.}
The  conformal weight of $G^\a(z)$ is $\D_\a=1+\frac{\a^2}{2g}$ and,
in the case $\bg=su(2)$,  $G^\a$'s  reduce to the two supercurrents of
the $N=2$ minimal model.
We also consider the $u(1)^n$ currents defined by
\eqn\uone{J^a(z)=\frac{i}{\a_+}\der\vp_a(z)=\sum_{n\in\bZ} z^{-n-1}J^a_n,\quad
 a=1,\ldots,n\, .}
The operators $G^\a$, $J$ and $T_\a$ are the basic generators of
the chiral algebra and since parafermions can be represented by a set of
bosons
as mentioned in the previous section, one can investigate the
structure of the chiral algebra without ambiguity starting from $G^\a$, $J$
and
$T_\a$.
It turns out that our chiral algebra  is   non-local as well as non-linear.
Here we merely quote the OPEs we need in what follows:
\eqna\hsOPE
$$\eqalignno{
&G^\a(z)G^\b(z')=\frac{ \sqrt{ {\frac{c}{d}} }
K(\a,\b)}{(z-z')^{1-\a\cdot\b/g}}
\[G^{\a+\b}(z')+\cdots\]\,,&\hsOPE a\cr
\apar
&G^\a(z)G^{-\a}(z')=\frac{\frac{c}{d}}{(z-z')^{2\D_\a}}+
\frac{\inv{g}\a\cdot J(z')}{(z-z')^{2\D_\a-1}}\cr
\apar
&\hskip 2cm+\inv{(z-z')^{2\D_\a-2}}\[T_\a(z')+\inv{2g}\a\cdot\der J(z')
+\cdots\]\,.&\hsOPE b\cr}
$$

The explicit expressions  \Trep, \Grep\ and \uone\ imply
the existence of automorphism in the algebra:
\eqn\sflow{\eqalign{&T_\a(z)\->T_\a(z)+
\frac{i}{g}\(\a\cdot\frac{d\b(z)}{dz}\)(\a\cdot J(z))-\frac{c}{2d}
\(\a\cdot\frac{d\b(z)}{dz}\)^2\,,\cr
\apar
&J(z)\->J(z)+i\frac{cg}{d}\frac{d\b(z)}{dz}\,,\cr
\apar
&G^\a(z)\->e^{i\a\cdot\b(z)}G^\a(z)\,,\cr}}
generated by the
$u(1)^n$ gauge transformation
\eqn\?{\vp(z)\->\vp(z)+\inv{\a_+}\b(z)\,,}
where $\b(z)$ is an $n$-dimensional vector whose components are
 arbitrary analytic functions.
Setting in  particular $\b(z)=i\eb\log z$ in \sflow\ with $\eb$ being an
$n$-dimensional vector we have
\eqn\sflowi{\eqalign{
       &L_m^\a\->L_m^\a-\inv{g}(\a\cdot\eb)(\a\cdot J_m)+
         \frac{c}{2d}(\a\cdot\eb)^2\,\d_{m,0}\, ,\cr
\apar
       &J_m^a \->J_m^a-\frac{cg}{d}\eb^a\,\d_{m,0}\, ,\cr
\apar
      &G^\a_r\->G^\a_{r+\a\cdot\eb}\, ,\cr}}
where
\eqn\susymode{G^\a(z)=\sum_r z^{-r-\D_\a}\,G_r^\a\,.}
This is a natural generalization of the one-parameter spectral flow in
$N=2$ superconformal algebra \rSS\ to the $n$-parameter flow in the
higher-rank
 case.

\subsec{The NS sector and Chiral Primary Fields}

We next seek for the analogue of the chirality condition and chiral primary
fields in $N=2$ superconformal field theory \rLVW.
 For this purpose let us first define the fields in the NS sector as the
set of fields $A_{NS}(z)$ which are local with respect to $G^\a(z)\,$:
\eqn\?{G^\a(z)A_{NS}(0)=\sum_{r\in\bZ-\D_\a}z^{-r-\D_\a}(G_r^\a A_{NS})(0)\,.}
If a state $\ket{\P_p}=\P_p(0)\ket{0}$, where  $\P_p(z)$ is
a field in  the NS sector, is annihilated by all
the positive-mode generators $L_m$, $J_m^a$ and $G^\a_{m-\frac{\a^2}{2g}}$
with $m>0$, then it is called a {\it primary} state.
A primary state $\ket{\P_p}$ is labeled by the eigenvalues of
 $L_0$ and  $J_0^a$
\eqn\?{L_0\ket{\P_p}=h\ket{\P_p}\,,\quad J_0^a\ket{\P_p}=q_a\ket{\P_p}\,,
\quad a=1,\ldots,n\,.}
We also define a {\it chiral\/} state $\ket{\P_c}$ in the NS sector by
requiring
\eqn\chiralcond{G_{-\frac{{\a_i}^2}{2g}}^{\a_i}\ket{\P_c}=0\,,
\quad i=1,\ldots,n\,,}
where we recall $\a_1,\ldots,\a_n$ are the simple roots of $\bg$.

 If a chiral state also satisfies the
primary state conditions it is called a {\it chiral primary\/} state.
Actually we have
\eqn\cpc{
G_{l-\frac{\a^2}{2g}}^{\a}\ket{\P_{c\wedge p}}=0\,,\quad l\ge 0\,,
\quad \a\in\D_+\,,}
for any chiral primary state $\ket{\P_{c\wedge p}}$.
This can be seen as follows.
First it is clear that \cpc\ is satisfied when $\a$ is a simple root.
By resorting to the standard procedure \refs{\rZFi,\rZFiii} one can deduce
from \hsOPE{a}\ that in the NS sector
\eqnn\GaGb
$$\eqalignno{\sum_{l=0}^\infty C_l^{(-\a\cdot\b/g)}&\(
G^\a_{p-l-\frac{\a\cdot(\a+2\b)}{2g}}\,G^\b_{q+l-\frac{\b^2}{2g}}+
G^\b_{q-l-\frac{\b\cdot(\b+2\a)}{2g}}\,G^\a_{p+l-\frac{\a^2}{2g}}\)\cr
\apar
&=\sqrt{\frac{c}{d}}\,K(\a,\b)\,G^{\a+\b}_{p+q-\frac{(\a+\b)^2}{2g}}\,,
&\GaGb\cr}$$
where $p,q\in\bZ$ and
\eqn\?{C_l^{(a)}=(-1)^l\frac{\G(a+1)}{l!\,\G(a+1-l)}\,.}
Suppose that \cpc\ is true for two positive roots $\a$ and $\b$ such that
$\a+\b$ is again a positive root. Then having  both sides of \GaGb\
with  $p=q=0$ act on the
chiral primary state $\ket{\P_{c\wedge p}}$  we find
\eqn\?{G^{\a+\b}_{-\frac{(\a+\b)^2}{2g}}\ket{\P_{c\wedge p}}=0\,.}
Hence taking account of the primary conditions,  \cpc\ is also true for the
positive root $\a+\b$. This completes the proof.

It also follows from \hsOPE{b}\ that
in the NS sector
\eqnn\GaGa
$$\eqalignno{
\sum_{l=0}^\infty C_l^{(\a^2/g-1)}&\(
G_{m-l+\frac{\a^2}{2g}}^{-\a}
\,G_{m'+l-\frac{\a^2}{2g}}^\a+G_{m'-1-l+\frac{\a^2}{2g}}^\a
\,G_{m+1+l-\frac{\a^2}{2g}}^{-\a}\)  \cr
\apar
&=\[ \frac{c}{2d}m(m+1)\d_{m+m',0}+L_{m+m'}^\a-\inv{2g}(m-m'+1)\a\cdot
J_{m+m'} \]\,,\cr
\apar
&&\GaGa\cr}
$$
where $m$, $m'\in\bZ$.
Let us consider a secondary state $G_{-\frac{\a^2}{2g}}^\a\ket{\P_p}$
for a NS primary state  $\ket{\P_p}$ and  $\a\in \D_+$.
 Define the adjoint of $G_r^\a$ by $(G_r^\a)^\dagger=G_{-r}^{-\a}$.
Using \GaGa\  we evaluate the norm of this secondary state.
It is then immediate to find a  bound imposed by unitarity
\eqn\?{h_\a\ge \inv{2g}\a\cdot q\,,}
where $L_0^\a\ket{\P_p}=h_\a\ket{\P_p}$.
For a  chiral primary state this bound is saturated and
hence there holds a linear relation between its conformal weight $h$ and
$u(1)^n$ charge $q$ :
\eqn\chirallin{h=\inv{g}\r\cdot q\,,}
Having  $G_{-1-\frac{{\a}^2}{2g}}^{\a}$ with $\a\in\D_+$ act on
 a chiral primary state and evaluating its norm we get a unitarity bound
\eqn\?{\inv{g}\a\cdot q\le\frac{c}{d}\,,\quad \a\in \D_+\,,}
and hence we find the restriction
\eqn\?{0\le h\le\frac{d-n}{4d} c\,,}
for chiral primary states.
As a result the number of chiral primary fields
is finite as long as the theory is non-degenerate.

The explicit form of chiral primary fields
is obtained as
\eqn\?{\left\{\P_\La(z)=\Ps^{\La,0}_\La(z)\ \big\vert\ \La\in
P_+^k\right\}\,.}
Using \GaGa\ and \eigenpf\ it is easy to check that the conditions \cpc\
are fulfilled for
$\ket{\P_\La}$
and  the relation \chirallin\ is satisfied:
\eqn\?{h_\La=\inv{g}\r\cdot q_\La\quad\hbox{with}\quad
q_\La=\frac{g}{k+g}\La\,.}
Then following the standard argument we find that there are no short-distance
singularities between two chiral primary fields
\eqn\cOPE{\P_\La(z)\P_{\La'}(z')\sim\P_{\La+\La'}(z')\, ,}
where we should understand that $\P_{\La+\La'}(z)=0$ if
$\La+\La'\not\in P_+^k$. Hence a finite nilpotent  ring is generated by
 our chiral primary fields. We will refer to  this ring as  {\it chiral ring}.

One can more generically  consider a set of conditions
\eqn\chiralcondii{G_{-\frac{{\a_i}^2}{2g}}^{w(\a_i)}\ket{\P}=0\,,
\quad i=1,\ldots,n\,,}
for each $w\in W$ instead of \chiralcond.
Denote as $\CR_w$ the set of primary fields in the NS sector obeying
\chiralcondii.  Then we have
\eqn\?{\CR_w=\left\{\Ps^{\La,0}_{w(\La)}\ \big\vert\ \La\in P_+^k\right\}\,.}
%
\insertfig{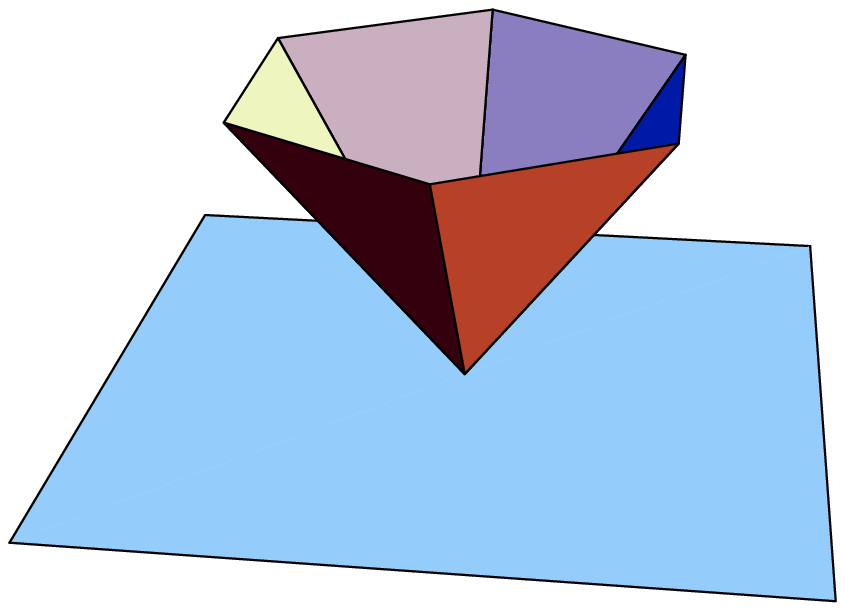}{5cm}
{The distribution of $(h,q)$ for $\mathop{\cup}\limits_{w\in W}\CR_w$
in the case $\bg=su(3)$. The $h$-axis is chosen to be vertical.
For easy reference the $h=0$ plane is also depicted. }
{\noindent Clearly}, in this notation $\CR_{\rm id}$ is the chiral ring
mentioned above.
As in the WZW model, the charge conjugate field of
$\Ps^{\La,0}_{w(\La)}\in\CR_w$ is
$\Ps^{\bar\La,0}_{ww_0(\bar\La)}\in\CR_{ww_0}$.
Therefore, in view of the property
$\La\in P_+^k\Leftrightarrow \bar\La\in P_+^k$,
the conjugate set of $\CR_w$ coincides with  $\CR_{ww_0}$.
Each $\CR_w$, $w\in W$ becomes  a ring isomorphic to the chiral ring
$\CR_{\rm id}$. However it should be noted that although the set
$\mathop{\cup}\limits_{w\in W}\CR_w$ is apparently
conjugation symmetric, in general two fields $\p\in\CR_w$ and
$\p'\in\CR_{w'}$ with $w\ne w'$ are not mutually local and hence
 we cannot associate a good algebraic structure with
the whole set $\mathop{\cup}\limits_{w\in W}\CR_w$.

\subsec{The R sector}

 Specializing the spectral flow parameter $\bar\eta=\frac{\eta}{g}\rho$ with
$\eta \in {\bf R}$ in \sflow\  we obtain a one-parameter flow
\eqn\sflowi{\eqalign{
      L_m^\a&\->L_m^\a-\eta\frac{\a\cdot\r}{g^2}(\a\cdot J_m)+
       \eta^2\frac{c}{2d}
      \(\frac{\a\cdot\r}{g}\)^2\,\d_{m,0}\, ,\cr
\apar
       J_m^a&\->J_m^a-\eta\frac{c}{d}\,(\r)_a\,\d_{m,0}\, ,\cr
\apar
       G^\a_r&\->G^\a_{r+\eta\frac{\a\cdot\r}{g}}\, ,\cr}}
and hence
\eqn\?{L_m\->L_m-\eta\inv{g}\r\cdot J_m+\eta^2\frac{c}{24}\,\d_{m,0}\,.}
 In the sector obtained by applying this flow with $\eta=1$ to the NS sector
the
generators $G_{r}^{\a_1},\ldots,G_{r}^{\a_n}$ and $G_{r}^{-\theta}$
have integer modings.
 Thus this is a  sector analogous to one of the R sectors in $N=2$
superconformal theory.
 The chirality conditions \chiralcond\ as well as the primary state
conditions for $G_{r}^{-\theta}$ are converted into ground-state conditions
\eqn\?{G_r^{\a_i}\ket{R}=0\,, \quad G_r^{-\t}\ket{R}=0\,,
\quad r=0,1,2,\ldots}
It is clear that under this flow a chiral primary field
$\P_\La=\Ps^{\La,0}_\La$
turns into a Ramond ground field $\Ps^{\La,\r}_{\La+\r}$.
More generally, under the flow corresponding to the choice $\eb=\inv{g}w(\r)$,
$w\in W$ in \sflowi\
the set $\CR_w$ flows to the set
\eqn\?{\CV_w=\left\{\Ps^{\La,w(\r)}_{w(\La+\r)}(z)\ \big\vert\
\La\in P_+^k\right\}\,,}
by  the rule $\Ps^{\La,0}_{w(\La)}\->\Ps^{\La,w(\r)}_{w(\La+\r)}$.
An element $\Ps^{\La,w(\r)}_{w(\La+\r)}$ of $\CV_w$ has quantum numbers
\eqn\?{h=\frac{c}{24},\qquad
            q= \frac{g}{k+g}w(\La+\r)-w(\r)\,.}
This spectrum can also be obtained by an analysis of branching relations \rK.

The charge conjugate field of $\Ps^{\La,w(\r)}_{w(\La+\r)}\in\CV_w$ is
$\Ps^{\bar\La,ww_0(\r)}_{ww_0(\bar\La+\r)}\in\CV_{ww_0}$ and hence the
conjugate
set of $\CV_w$ is $\CV_{ww_0}$.
Therefore the whole set $\mathop{\cup}\limits_{w\in W}\CV_w$ of R ground
fields is symmetric under charge conjugation.
\insertfig{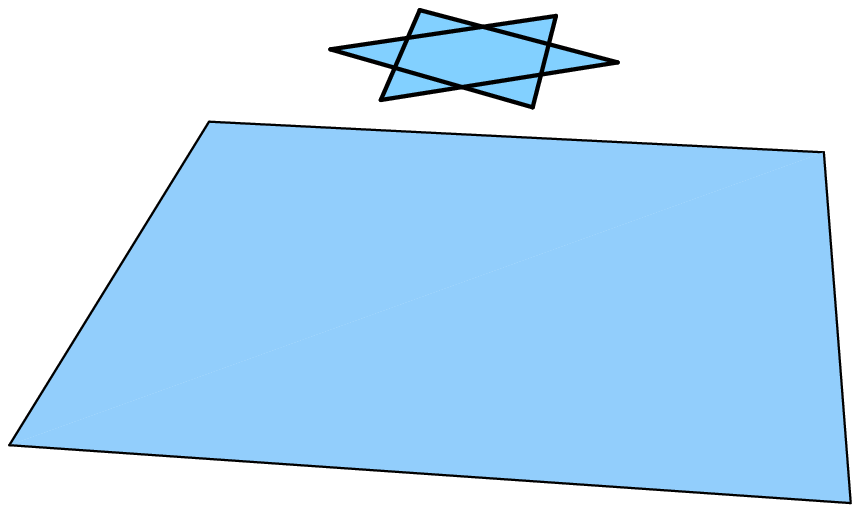}{5cm}
{The distribution of $(h,q)$ for $\mathop{\cup}\limits_{w\in W}\CV_w$ in the
case  $\bg=su(3)$ and the $h=0$ plane. `Star of David' sits at the height of
$c/24$.}
\insertfig{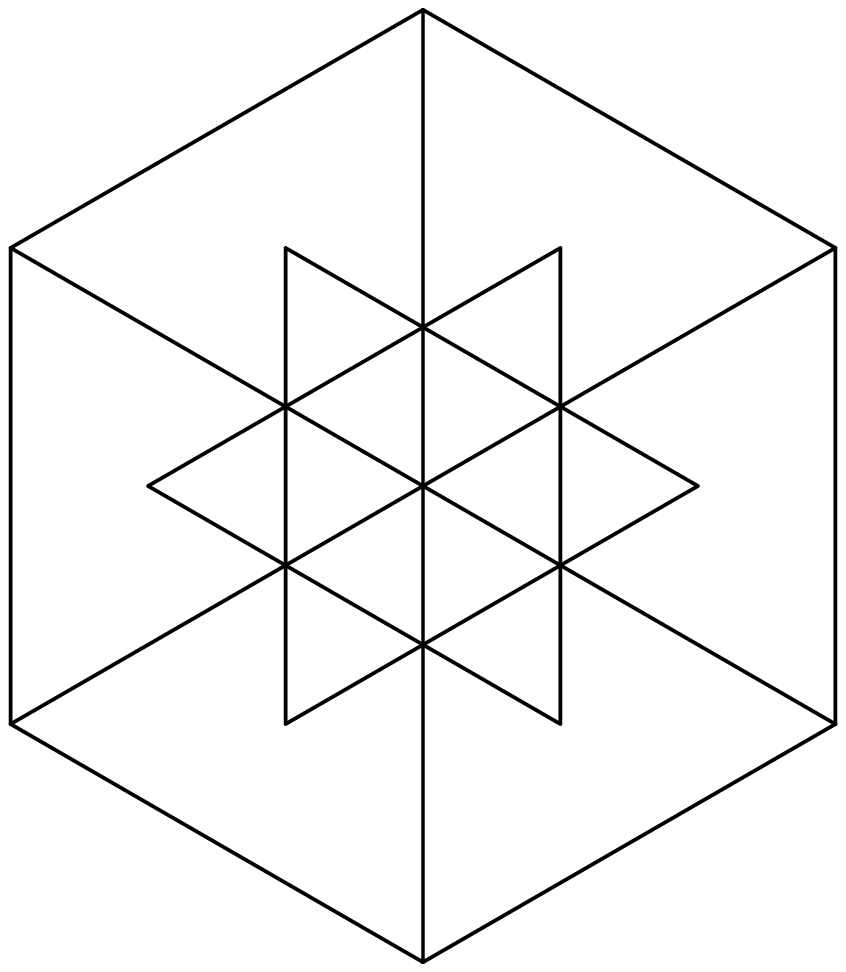}{5cm}
{The $u(1)^n$ charge  distribution of both $\mathop{\cup}\limits_{w\in
W}\CR_w$
and $\mathop{\cup}\limits_{w\in W}\CV_w$ for $\bg=su(3)$.
The six outer equilateral triangles forming a hexagon correspond to
$\mathop{\cup}\limits_{w\in W}\CR_w$ while
the two inside ones which are overlaid correspond to
$\mathop{\cup}\limits_{w\in W}\CV_w$.
Each outer triangle goes to one of the two inside ones by a spectral flow
mentioned in the text.}
At this juncture we should comment on the special feature of the case
$\bg=su(2)$, {\it i.e.\/} the \minimal. In this case we have $W=\{\pm 1\}$ and
$\CV_{w=+1}=\CV_{w=-1}$. The latter follows from
\eqn\sutwoconjsym{\eqalign{&\CV_{w=-1}\ni \p^l_{-l}(z)
\nord\exp\[+i\{\a_++\a_-(l+1)\}\vp(z)/\sqrt{2}\]\nord \cr
&\hskip 1.5cm=\p^{k-l}_{k-l}(z)
\nord\exp\[-i\{\a_++\a_-(k-l+1)\}\vp(z)/\sqrt{2}\]\nord
\in \CV_{w=+1}\,,\cr}}
where we wrote $\p^l_m(z)$ instead of $\p^{l/\sqrt{2}}_{m/\sqrt{2}}(z)$
following the standard convention.
In deriving \sutwoconjsym\ we used $g\a_+=-(k+g)\a_-$ and the symmetry
property
$\p^l_{-l}(z)=\p^{k-l}_{k-l}(z)$ which is the consequence of diagram
automorphism. Therefore in contrast to the generic case $\bg\ne su(2)$ where
$\CV_{w=\rm{id}}$ alone is certainly not charge-conjugation symmetric,
in the case $\bg=su(2)$ the set $\CV_{w=+1}$ is the charge-conjugation
symmetric set of the Ramond ground fields. This in turn implies that the set
of chiral primary fields $\CR_{w=+1}$, which is mapped onto
$\CV_{w=+1}$ by a spectral flow, also has a symmetry. This is nothing but the
Poincar{\' e} duality of the chiral ring.  We will say more about this in the
next section. Recall that it is quite generally
argued in $N=2$ superconformal field theory \rLVW\ that
$u(1)$ charge-conjugation symmetry of the Ramond vacua  translates into
the Poincar{\' e} duality of the chiral ring.

\newsec{Topological Conformal Field Theory}

As we have seen in the previous section the higher-rank supersymmetry
turns out to be a Lie algebraic extension of $N=2$ superconformal symmetry.
According to general arguments \refs{\rW,\rEY}\ every $N=2$ theory yields a
topological field theory.
So a natural problem arises of whether we can produce  a sort of
topological field theories starting from our higher-rank supersymmetric
models.
Answering this is the main theme of this section. We will argue
that it is possible to construct topological field theories but to render
 them sensible we must adopt some special procedures.

As in the twisted $N=2$ theory
we begin by assuming that the new stress-energy tensor is given by
\eqnn\twist
$$\eqalignno{{\hat T}(z)&=T(z)+\inv{g}\r\cdot\der J(z)\cr
\apar
&=-\ha\nord\der\vp(z)\cdot\der\vp(z)\nord+i\a_0\r\cdot\der^2\vp(z)+
T^{\rm(pf)}(z)\,.&\twist\cr}
$$
which generates the Virasoro algebra with a vanishing central charge.
In the usual twisted $N=2$ theory the twisted stress-energy tensor can be
expressed
as
\eqn\?{\hat T_{N=2}(z)= \oint_z dz'G^+(z')G^-(z)\,,}
and is claimed to be BRST-exact by interpreting $\oint G^+$ as the BRST
charge.
We want a similar `BRST-exact'\foot{
We will explain later what is meant by  `BRST-exact'.}
 expression for $\hat T(z)$.
To our luck there exists such a formula
\eqn\brstexact{{\hat T}(z)\propto T_s(z)\,,}
with $T_s(z)$ given by
\eqn\?{\eqalign{
T_s(z)&=\oint\, \prod_{i=1}^n\prod_{p=1}^{a_i} dz_{i,p}\,
G^{\a_1}(z_{1,1})\cdots G^{\a_1}(z_{1,a_1})\cr
\apar
&\hskip 12mm\times\cdots\times G^{\a_n}(z_{n,1})\cdots
G^{\a_n}(z_{n,a_n})G^{-\t}(z)\cr
\apar
&\equiv \oint (G^{\a_1})^{a_1}\cdots(G^{\a_n})^{a_n}G^{-\t}(z)\,,\cr}}
where the positive integers $\row a n$ are defined via
$\t=\sum_{i=1}^n a_i\a_i$ and the contour of the  multiple integrals
must be chosen appropriately\foot{Throughout the paper we are rather
cavalier about the treatment of
multiple integral contours although we do not pretend to have succeeded in
handling rigorously the contours.
Anyway simple-minded treatment indeed works as we shall  see.}.
Notice that $G^{\a_i}(z)$'s have conformal
weight 1 with respect to $\hat T(z)$ while $G^{-\t}(z)$ has 2.
The expression \brstexact\ was first proposed in \rEHY\ in the context of
$c=0$
GKO coset models.
We note that the twisted stress-energy tensor \twist\ coincides with
that appearing in the $G_k\times G_0/G_k$ model.
In what follows  we first dwell upon the validity of
\brstexact\
and then we move on to the actual construction of topological field theories.

\subsec{BRST-exactness of the Stress-Energy Tensor -- Multiple Integrals}

Our goal in this  subsection is to perform some sample
calculations explicitly  verifying  the BRST-exactness \brstexact\ of the
twisted stress-energy tensor.
It remains as a mathematically interesting (but perhaps
pains-taking) task to accomplish these calculations in full generality,
namely, to prove \brstexact\  for all simple Lie algebras with arbitrary
levels. We hope we can
return to this challenging problem in future.

Before launching to the calculations, we require some definitions; let
$M_{ij}$ and $m_i$'s be given by
\eqn\thstressaux{\eqalign{
            &M_{ij}=\frac{4}{\a_i^2\a_j^2}\ts\o_i\cdot\o_j\ts,\cr
\apar
            &2\r=\sum_{i=1}^n m_i\a_i\ts.\cr}}
In the sequel we will frequently use the  relation
\eqn\comprel{\sum_{i,j=1}^n M_{ij}(\a_i)_a(\a_j)_b=\d_{ab}\,.}
The necessary  Lie algebraic data  are
summarized in   {\bf Tables 1--3}.
\def\tphan{\vphantom{\hbox{\vrule height 1cm width 0mm depth -1cm}}}

\bigskip
\leftline{$\bullet$\ $\bg=\CA_n,\ k\ge 1$}
\smallskip
To evaluate $\hat T_s(z)$ it is convenient to employ the interacting boson
representation of parafermions reviewed in sect.2. After some standard
manipulations we find
\eqnn\aint
$$\eqalignno{\hat T_s(z)&\propto\oint\prod_{i=1}^n dz_i\cr
\apar
&\times\sum_{ j_0,\ldots,j_n=1}^k\,
\prod_{i=1}^{n-1}(z_i-z_{i+1})^{-(\d_{j_i,j_{i+1}}+\inv{g})}
    (z_1-z)^{-(\d_{j_1,j_0}+\inv{g})}(z_n-z)^{-(\d_{j_n,j_0}+\inv{g})}\cr
\apar
&\times\nord\exp\[i\a_+\sum_{i=1}^n\a_i\cdot(\vp(z_i)-\vp(z))\]\nord\cr
\apar
&\times\nord\exp\[i\sum_{i=1}^n\{\a_i\otimes\la_{j_i}\cdot\c(z_i)-
\a_i\otimes\la_{j_0}\cdot\c(z)\}\]\nord
  c_{\a_1}^{j_1}\cdots c_{\a_n}^{j_n} c_{-\t}^{j_0}\cr
\apar
&\propto\oint\prod_{i=1}^n dz_i\cr
\apar
&\times\[\prod_{i=1}^{n-1}(z_i-z_{i+1})(z_1-z)(z_n-z)\]^{-1-\inv{g}}
\Biggl[1+\ha\sum_{1\le i,j\le n} (z_i-z)(z_j-z)A_{ij}(z)\cr
\apar
&+\sum_{i=1}^n(z_i-z)^2B_i(z)
+(z_1-z)\sum_{i=1}^{n-1}(z_i-z_{i+1})\CT_{\a_1+\cdots+\a_i}(z)\cr
\apar
&+(z_1-z)(z_n-z)\CT_{\a_1+\cdots+\a_n}(z)
-\sum_{2\le i\le j \le
n-1}(z_{i-1}-z_i)(z_j-z_{j+1})\CT_{\a_i+\cdots+\a_j}(z)\cr
\apar
&-\sum_{2\le i \le n}(z_{i-1}-z_i)(z_n-z)\CT_{\a_i+\cdots+\a_n}(z)
+\hbox{higher order terms}\ \Biggr]\,,
\cr
\apar
&&\aint\cr}
$$
where
\eqn\thstressaux{\eqalign{
 &A_{ij}(z)=\nord(i\a_+\a_i\cdot\der\vp(z))
                    (i\a_+\a_j\cdot\der\vp(z))\nord
 +\inv{k}\nord(i\a_i\cdot\der\c(z))\cdot(i\a_j\cdot\der\c(z))\nord\ts,\cr
 &B_i(z)=\ha i\a_+\a_i\cdot\der^2\vp(z)\ts,\cr}}
and
\eqn\pfexp{
\CT_\a(z)=\inv{k}\sum_{\b\in\D(su(k))}
\nord\exp[i\a\otimes\b\cdot\c(z)]\nord\,c^\b_\a\,,
\quad \a\in\D_+\,.}
In deriving \aint\ we have used the properties of cocycle factors summarized
in
sect.2.

Next  we  make change of variables
$z_i\->z+\prod_{l=1}^iu_l$, $i=1,\ldots,n$ and integrate  over $u_1$
by the Cauchy residue formula to  reach
\eqnn\aintii
$$\eqalignno{
\hat T_s(z)&\propto\oint\prod_{i=2}^n\, du_i\,
              u_i^{\(\frac{i}{g}-1\)-1}(1-u_i)^{-\inv{g}-1}\cr
\apar
&\times\Biggr[\ha\(A_{11}(z)+\sum_{i=1}^n(u_2\cdots u_i)^2\,A_{ii}(z)\)
+\sum_{i=2}^n (u_2\cdots u_i)\, A_{1i}(z)\cr
\apar
&+\sum_{2\le i<j\le n}(u_2\cdots u_i)(u_2\cdots u_j)\,A_{ij}(z)
+B_1(z)+\sum_{i=2}^n(u_2\cdots u_i)^2\,B_i(z)\cr
\apar
&+(1-u_2)\,\CT_{\a_1}(z)+\sum_{i=2}^{n-1}(u_2\cdots u_i)(1-u_{i+1})
\,\CT_{\a_1+\cdots+\a_i}(z)\cr
\apar
&+(u_2\cdots u_n)\,\CT_{\a_1+\cdots+\a_n}(z)
-u_2(1-u_2)(1-u_3)\,\CT_{\a_2}(z)\cr
\apar
&-\sum_{i=3}^{n-1}(u_2\cdots u_{i-1})^2
u_i(1-u_i)(1-u_{i+1})\,\CT_{\a_i}(z)\cr
\apar
&-\sum_{i=3}^{n-1}u_2(1-u_2)u_3\cdots u_i(1-u_{i+1})\,
\CT_{\a_2+\cdots+\a_i}(z)\cr
\apar
&-\sum_{3\le i<j\le n-1}(u_2\cdots u_{i-1})^2 u_i(1-u_i)u_{i+1}\cdots u_j
(1-u_{j+1})\,\CT_{\a_i+\cdots+\a_j}(z)\cr
\apar
&-u_2(1-u_2)(u_3\cdots u_n)\,\CT_{\a_2+\cdots+\a_n}(z)\cr
\apar
&-\sum_{i=3}^{n-1}(u_2\cdots u_{i-1})^2 u_i(1-u_i)(u_{i+1}\cdots u_n)\,
\CT_{\a_i+\cdots+\a_n}(z)\cr
&-(u_2\cdots u_{n-1})^2 u_n(1-u_n)\CT_{\a_n}(z)
\Biggl]\,.&\aintii\cr}$$

By repeatedly applying  the recursion formulae of the beta functions
it is now straightforward to check that
\eqn\atypefin{
\hat T_s(z)\propto \frac{g}{2}\sum_{i,j=1}^n M_{ij}A_{ij}(z)
+\sum_{i=1}^n m_iB_i(z)+\sum_{\a\in\D_+}\, \CT_\a(z)\,,}
which amounts to \brstexact\ with $T^{\rm(pf)}(z)$ given by \pfstboson.
\bigskip
\leftline{$\bullet$\ $\bg=\CB_n,\ k=1$}
\smallskip
In this case the parafermions are given by
\eqn\bpara{\ps_\a(z)=\cases{\ps(z)\,,&\hbox{if $\a$ is short}, \cr
\apar
1\,,&\hbox{if $\a$ is long},\cr}}
where $\ps(z)$ is a Majorana fermion satisfying the OPE
\eqn\mafermi{\ps(z)\ps(z')=\inv{z-z'}+\cdots\,,}
Consequently,
\eqn\bparasti{T_\a^{\rm(pf)}(z)=\cases{\nord\ps'(z)\ps(z)\nord\,,&
\hbox{if $\a$ is short},\cr
\apar
0\,,&\hbox{if $\a$ is long},\cr}}
and hence
\eqn\bparastii{T^{\rm (pf)}(z)=\inv{1+g}
\sum_{\a\in \D_+}T_\a^{\rm(pf)}(z)=\ha \nord\ps'(z)\ps(z)\nord\,.}

Again it is easy to see
\eqnn\btype
$$\eqalignno{
\hat T_s(z)&\propto \oint\,dz_1\,\prod_{i=2}^n\ts dz_{i,1}dz_{i,2}\cr
\apar
&\times\[\prod_{p=1}^2(z_1-z_{2,p})^{-1}(z_{2,p}-z)^{-1}
\prod_{i=2}^{n-1}\ts\prod_{p,q=1}^2\ts(z_{i,p}-z_{i+1,q})^{-1}\,
(z_{i,1}-z_{i,2})^2\]^{1+\inv{g}}\cr
\apar
&\times(z_{n,1}-z_{n,2})^{\inv{g}}
\nord\exp\[i\a_+\Biggl\{\a_1\cdot(\vp(z_1)-\vp(z))
+\sum_{i=2}^n\sum_{p=1}^2\a_i\cdot(\vp(z_{i,p})-\vp(z))\Biggr\}\]\nord\cr
\apar
&\times\[1+(z_{n,1}-z_{n,2})\nord\ps(z_{n,1})\ps(z_{n,2})\nord\]\ts.
&\btype\cr}$$
Setting
\eqn\ABC{\eqalign{&A_{ij}(z)=\nord(i\a_+\a_i\cdot\der\vp(z))
                        (i\a_+\a_j\cdot\der\vp(z))\nord\ts,\cr
\apar
&B_i(z)=\ha i\a_+\a_i\cdot\der^2\vp(z)\ts,\cr
\apar
&C(z)=\nord\ps'(z)\ps(z)\nord\ts,\cr}}
we obtain
\eqnn\lemmab
\proclaim Lemma \lemmab.
 For each $l=3,4,\ldots,n$,
\eqnn\bint
$$\eqalignno{\hat T_s(z)&\propto
           \oint\ts \prod_{i=l}^n\ts dz_{i,1}dz_{i,2}\cr
\apar
&\times\prod_{p=1}^2\ts\( z_{l,p}(1-z_{l,p}) \)^{-1-\inv{g}}\ts
(z_{l,1}-z_{l,2})^{-1+\frac{2}{g}l}\,P_l^{\CB}(z_{\bullet,\bullet})\cr
\apar
&\times\Biggl[
 \Biggl(\frac{g}{2} \sum_{\scriptstyle 1\le i,j \le n\atop
\scriptstyle \min(i,j)\le l-1}\!
 M_{ij}\ts A_{ij}(z)+\sum_{i=1}^{l-1}m_i\ts B_i(z) \Biggr)
 +2\sum_{l\le i,j\le n} f_l(\z_i,\z_j)\ts A_{ij}(z)\cr
&\qquad+\sum_{i=l}^n\ts \sum_{p=1}^2\ts f_l(z_{i,p},z_{i,p})\ts B_i(z)
 +d_l(z_{n,1}-z_{n,2})^2 \ts C(z)
 \Biggr]&\bint\cr
}$$
where we have set $\z_i=\ha(z_{i,1}+z_{i,2})$ for $i=3,4,\ldots,n$ and
\eqnn\?
$$\vbox{\eqalignno{
P_l^{\CB}&(z_{\bullet,\bullet})=\cr
\apar
&=\cases{\displaystyle
\Biggl[\prod_{p=1}^2 z_{l+1,p}^{-1}(1-z_{l+1,p})^{-1}\cr
\apar
\displaystyle\ \ \times\prod_{i=l+1}^{n-1}\ts\prod_{p,q=1}^2\ts
 (z_{i,p}-z_{i+1,q})^{-1}\ts
(z_{i,1}-z_{i,2})^2\Biggr]^{1+\inv{g}}(z_{n,1}-z_{n,2})^\inv{g}\,,&\cr
\apar
\hskip 5cm &$3\le l\le n-2\,,$\cr
\apar
\displaystyle\[\prod_{p=1}^2z_{n,p}^{-1}(1-z_{n,p})^{-1}\]^{1+\inv{g}}
(z_{n,1}-z_{n,2})^\inv{g}\,,&$l=n-1\,,$\cr
\apar
\displaystyle 1\,,&$l=n\,.$\cr}\cr
\apar
&&\?\cr}}
$$
The functions $f_l(u,v)$ in \bint\ are defined for $l\ge 3$ by
\eqn\bdaux{f_l(u,v)=a_l(uv+(1-u)(1-v))+b_l(u(1-v)+v(1-u))+c_l\,,}
with
\eqn\recsolabc{\eqalign{a_l&=\ha g(l-1)-(l-2)^2\ts,\cr
                     b_l&=l-2\ts,\cr
                     c_l&=\ha(l-2)(l-3)\ts,\cr}}
and
\eqn\recsold{d_l=(l-1)(g-2(l-2))\,.}

{\noindent {\bf Proof.} \hskip 5mm
For $3\le l \le n-2$ we prove this by induction on $l$.\hfil
\medskip
\item{(i)}$l=3$. \hfill\break
First make change of variables
$z_{2,1}\->z+x$, $z_{2,2}\->z+xy$, $z_1\->z+xz_1$ and
$z_{i,p}\->z+xz_{i,p}$ for $i=3,\ldots,n$ and $p=1,2$ on the RHS
 of \btype\ and then integrate over
$x$. Make further replacement $z_1\->z_1+(1-z_1)y$ and
$z_{i,p}\->z_{i,p}+(1-z_{i,p})y$ for $i=3,\ldots,n$ and $p=1,2$,
then integrate out $z_1$ and $y$ using the recursion relations
of the beta functions. Finally after performing change of variables
$z_{i,p}\->z_{3,1}z_{i,p}+z_{3,2}(1-z_{i,p})$ for $4\le i\le n$ and
$p=1,2$ we arrive at \bint\ with $l=3$.
\smallskip
\item{(ii)}$\hbox{\it The inductive step}\ l\->l+1$.\hfill\break
Suppose \bint\ is true for an $l$ such that $3\le l< n-2$ and set
$z_{l,1}=x$ and $z_{l,2}=y$. Let  $\dvev{f}_l=\vev{f}_l/\vev{1}_l$ where
\eqn\btypesel{\vev{f}_l=\oint dxdy\, f(x,y) \[x(1-x)y(1-y)\]^{-1-\inv{g}}
(x-y)^{-1+\frac{2}{g}l}\,.}
We know from the results in appendix  that
\eqn\twosel{\eqalign{&\dvev{x+y}_l=1\,,\qquad \dvev{xy}_l=\inv{g-2(l-2)}\,,\cr
\apar
&\dvev{(x-y)^2}_l=\frac{l}{l-1}\ts
                  \frac{g-2(l-1)}{g-2(l-2)}\,.\cr}}
To prove \bint\ for $l+1$ it suffices to show that
\eqnn\auxp\eqna\auxprops\eqnn\recursiond
$$\eqalignno{&\dvev{{f_l(xu+y(1-u),xv+y(1-v))}}_l=f_{l+1}(u,v)\,,&\auxp\cr
\apar
&\ddvev{f_l\(\frac{x+y}{2},\frac{x+y}{2}\)}_l=\frac{gl}{4}\ts,&\auxprops a\cr
\apar
&\ddvev{f_l\(\frac{x+y}{2},xv+y(1-v)\)}_l=\frac{gl}{4}\ts,&\auxprops b\cr
\apar
&\dvev{f_l(x,x)+f_l(y,y)}_l=l(g+1-l)\ts,&\auxprops c\cr
\apar
&d_{l+1}=d_l\dvev{(x-y)^2}_l\,,&\recursiond\cr}
$$
because then we can reach the desired result first by   integrating over $x$
and $y$  while  taking into account the data in
{\bf Table 3} and then by  performing  change of variables
$z_{i,p}\->z_{l+1,1}z_{i,p}+z_{l+1,2}(1-z_{i,p})$ for $l+2\le i \le n$ and
$p=1,2$.
The proof of \auxp--\recursiond\ goes as follows.
One finds from \twosel\ that eqs.\auxp\ and \recursiond\
are equivalent to the recursion relations
\eqnn\recursion
$$\eqalignno{a_{l+1}&=\frac{lg-2(l-1)^2}{(l-1)(g-2(l-2))}(a_l-b_l)\ts,\cr
\apar
b_{l+1}&=\inv{g-2(l-2)}(a_l-b_l)\ts,\cr
\apar
c_{l+1}&=b_l+c_l\ts,\cr
\apar
d_{l+1}&=\frac{l}{l-1}\frac{g-2(l-1)}{g-2(l-2)}d_l\ts,
                                          &\recursion\cr}
$$
whose  solutions  with the initial conditions
$a_3= g-1$, $b_3=1$, $c_3=0$ and $d_3=2g-4$ can easily be shown to be given by
\recsolabc\ and \recsold.  By substituting \recsolabc\ and \recsold\
into \bdaux\ and using \twosel, we obtain \auxprops{a\hbox{--}c}.
\medskip
Now that \bint\ is true for $l=n-2$ the case  $l=n-1$ and then the case $l=n$
can be proven by similar calculations as above.
\qed
\bigskip
If we take $l=n$ in \bint, the remaining integrations over $z_{n,1}$ and
$z_{n,2}$ can be readily done via the formulas appearing in the
above proof to find that
\eqn\gfinal{\hat T_s(z)\propto
\frac{g}{2}\sum_{i,j=1}^n M_{ij}A_{ij}(z)
+\sum_{i=1}^n m_iB_i(z)+d_{n+1}\,C(z)\,,}
which immediately yields \brstexact.
\bigskip
\leftline{$\bullet$\ $\bg=\CD_n,\ k=1$}
\smallskip
In this case we need not introduce parafermions and we have
\eqnn\dtype
$$\eqalignno{
\hat T_s(z)&\propto \oint\,
dz_1\,\prod_{i=2}^{n-2}\ts dz_{i,1}dz_{i,2}\,dz_{n-1}dz_n\cr
\apar
&\times\Biggl[\prod_{p=1}^2(z_1-z_{2,p})^{-1}(z_{2,p}-z)^{-1}
\prod_{i=2}^{n-3}\ts\prod_{p,q=1}^2\ts(z_{i,p}-z_{i+1,q})^{-1}\cr
\apar
&\quad\times\prod_{p=1}^2(z_{n-2,p}-z_{n-1})^{-1}(z_{n-2,p}-z_n)^{-1}
\prod_{i=2}^{n-2}\ts(z_{i,1}-z_{i,2})^2\Biggr]^{1+\inv{g}}\cr
\apar
&\quad\times\nord\exp\Biggl[
i\a_+\Biggl\{\sum_{\m=1,n-1,n}\a_\m\cdot(\vp(z_\m)-\vp(z))\cr
\apar
&\hskip 3.5cm+\sum_{i=2}^{n-2}\sum_{p=1}^2\a_i\cdot(\vp(z_{i,p})-
\vp(z))\Biggr\}\Biggl]
\nord\,.
&\dtype\cr}$$

Let $A_{ij}(z)$ and $B_i(z)$ be as in \ABC\ and
set $\z_i=\ha(z_{i,1}+z_{i,2})$ for $i=3,\ldots,n-2$.
Let $f_l(u,v)$'s be defined as in {\bf Lemma \lemmab}.
Quite a similar reasoning as in the previous case entails
\eqnn\lemmad
\proclaim Lemma \lemmad.
For each $l=3,4,\ldots,n-3$,
\eqnn\dint
$$\eqalignno{\hat T_s(z)&\propto
           \oint\ts\prod_{i=l}^{n-2}\ts dz_{i,1}dz_{i,2}dz_{n-1}dz_n
            \ts\prod_{p=1}^2\ts
         \( z_{l,p}(1-z_{l,p}) \)^{-1-\inv{g}}\ts
         (z_{l,1}-z_{l,2})^{-1+\frac{2}{g}l}\cr
\apar
&\times P_l^{\CD}(z_{\bullet,\bullet})
\prod_{\mu=n-1,n}\ts \prod_{p=1}^2(z_{n-2,p}-z_\mu)^{-1-\inv{g}}\cr
\apar
&\times\Biggl[
 \Biggl(\frac{g}{2} \sum_{\scriptstyle 1\le i,j\le n \atop
\scriptstyle \min (i,j)\le l-1}\!
 M_{ij}\ts A_{ij}(z)+\sum_{i=1}^{l-1}m_i\ts B_i(z) \Biggr)
+2\sum_{l\le i,j\le n-2} f_l(\z_i,\z_j)\ts A_{ij}(z)\cr
\apar
&+2\sum_{\mu=n-1,n}\ts \sum_{i=l}^{n-2}\ts f_l(z_\mu,\z_i)\ts A_{i\mu}(z)
+\ha\sum_{\mu=n-1,n}\ts f_l(z_\mu,z_\mu)\ts(A_{\mu\mu}+2B_\mu)(z)\cr
\apar
&+f_l(z_{n-1},z_n)\ts A_{n-1,n}(z)
+\sum_{i=l}^{n-2}\ts\sum_{p=1}^2\ts
f_l(z_{i,p},z_{i,p})\ts B_i(z)\Biggr]\,,&\dint\cr
}$$
where
\eqnn\?
$$\vbox{\eqalignno{&P_l^{\CD}(z_{\bullet,\bullet})=\cases{
\displaystyle\Biggl[\prod_{p=1}^2z_{l+1,p}^{-1}(1-z_{l+1,p})^{-1}
\prod_{i=l+1}^{n-3}\ts\prod_{p,q=1}^2\ts(z_{i,p}-z_{i+1,q})^{-1}\cr
\apar
\displaystyle\hskip 5cm\times\prod_{i=l+1}^{n-2}\ts
(z_{i,1}-z_{i,2})^2\Biggr]^{1+\inv{g}}\,,&$3\le l \le n-4\,,$\cr
\apar
\displaystyle\[\prod_{p=1}^2 z_{n-2,p}^{-1}(1-z_{n-2,p})^{-1}
(z_{n-2,1}-z_{n-2,2})^2\]^{1+\inv{g}}\,,&$l=n-3\,,$\cr}\cr
\apar
&&\?\cr}}
$$

Now set $l=n-3$ in this Lemma, integrate out $z_{n-3,1}$ and $z_{n-3,2}$
and  make change of variables
$z_\m\->z_{n-2,1}z_\m+z_{n-2,2}(1-z_\m)$ for $\m=n-1, n$.
Then  further integrating over $z_{n-2,1}$ and
$z_{n-2,2}$, we find
\eqnn\dintiii
$$\eqalignno{\hat T_s(z)&\propto\oint\,dz_{n-1}dz_n\,
\[z_{n-1}(1-z_{n-1})z_n(1-z_n)\]^{-1-\inv{g}}\cr
\apar
&\times\Biggl[
 \Biggl(\frac{g}{2} \sum_{\scriptstyle 1\le i,j \le n-2\atop
\scriptstyle \min (i,j)\le n-2}\!
 M_{ij}\ts A_{ij}(z)+\sum_{i=1}^{n-2}m_i\ts B_i(z) \Biggr)\cr
\apar
&\quad+\ha\sum_{\mu=n-1,n}\ts f_{n-1}(z_\mu,z_\mu)\ts(A_{\mu\mu}+2B_\mu)(z)+
f_{n-1}(z_{n-1},z_n)\ts A_{n-1,n}(z)\Biggr]\,.\cr
\apar
&&\dintiii\cr}$$
After performing the remaining  integrations over $z_{n-1}$ and $z_n$,
which can be  done
without difficulty, one ends up with
\eqn\dtypefin{\eqalign{
\hat T_s(z)&\propto \frac{g}{2}\sum_{i,j=1}^n M_{ij}A_{ij}(z)
+\sum_{i=1}^n m_iB_i(z)\,,}}
which is equivalent to \brstexact.
\bigskip
\leftline{$\bullet$\ $\CG_2$, $k=1$}
\smallskip
The parafermions in this case can be represented as
\eqn\gpara{\ps_\a(z)=\cases{\displaystyle\inv{\sqrt{3}} \sum_{j=1}^3
\nord \exp\[ i\sqrt{2}\la_j\cdot X(z)\]\nord\,,&\hbox{if $\a$ is short,}\cr
\apar
1\,,&\hbox{if $\a$ is  long,}\cr}}
where $\la_1$, $\la_2$ and $\la_3$ are the weights in the vector
representation of $su(3)$
and $X(z)=(X_1(z),X_2(z))$ are free bosons.
It is easy to see that if $\a$ is a short root
\eqn\gstensori{
T_\a^{\rm (pf)}(z)=
\inv{3}\[-\nord\der X(z)\cdot\der X(z)\nord+\sum_{\b\in\D(su(3))}
                \nord\exp[i\sqrt{2}\b\cdot X(z)]\nord\]\,,}
 and otherwise $T_\a^{\rm (pf)}(z)=0$,  so that
\eqn\gstensorii{T^{\rm (pf)}(z)=\frac{c}{d}\[
-\nord\der X(z)\cdot\der X(z)\nord+\sum_{\b\in\D(su(3))}
                \nord\exp[i\sqrt{2}\b\cdot X(z)]\nord\]\,.}

This representation of parafermions immediately gives
\eqnn\gtype
$$\eqalignno{
\hat T_s(z)&\propto \oint\,
\prod_{p=1}^3dz_{1,p}\,\prod_{q=1}^2dz_{2,q}\cr
\apar
&\times\[\prod_{p=1}^3\prod_{q=1}^2
(z_{1,p}-z_{2,q})^{-1}\prod_{q=1}^2(z_{2,q}-z)^{-1}
(z_{2,1}-z_{2,2})^2\]^{1+\inv{g}}\cr
\apar
&\times\sum_{j_1,j_2,j_3=1}^3\prod_{1\le p<q\le 3}
(z_{1,p}-z_{1,q})^{2(\d_{j_p,j_q}+\inv{3g})}
\nord\exp\Biggl[i\a_+\Biggl\{\sum_{p=1}^3\a_1\cdot(\vp(z_{1,p})-\vp(z))\cr
\apar
&+\sum_{q=1}^2\a_2\cdot(\vp(z_{2,q})-\vp(z))\Biggr\}\Biggr]\nord\,
\nord\exp\[i\sqrt{2}\sum_{p=1}^3\la_{j_p}\cdot\vp(z_{1,p})\]\nord\,.
&\gtype\cr}
$$

We first make change of variables
$z_{2,1}\->z+x$, $z_{2,2}\->z+xy$, $z_{1,p}\->z+xz_{1,p}$\  $(p=1,2,3)$, and
then integrate over $x$. Further replacement
$z_{1,p}\->z_{1,p}+(1-z_{1,p})y$\  for $p=1,2,3$ yields
\eqnn\gint
$$\eqalignno{
\hat T_s(z)&\propto \oint\,dy \prod_{p=1}^3dz_{1,p}\,
y^{-\frac{5}{4}}
(1-y)^{-\frac{3}{2}}\prod_{p=1}^3 z_{1,p}^{-\frac{5}{4}}
(1-z_{1,p})^{-\frac{5}{4}}
\prod_{1\le p<q\le 3}(z_{1,p}-z_{1,q})^\inv{6}\cr
\apar
&\times\Biggl[\ha \(\sum_{p=1}^3\{z_{1,p}+(1-z_{1,p})y\}\)^2\, A_{11}(z)\cr
\apar
&+(1+y)\sum_{p=1}^3\{z_{1,p}+(1-z_{1,p})y\}\,A_{12}(z)
+\ha(1+y)^2\,A_{22}(z)\cr
\apar
&+\sum_{p=1}^3\{z_{1,p}+(1-z_{1,p})y\}^2\, B_1(z)+
(1+y^2)\,B_2(z)\cr
\apar
&+\inv{6}(1-y)^2\sum_{1\le p<q\le 3}(z_{1,p}-z_{1,q})^2\,
C(z)\Biggr]\,,&\gint\cr}$$
where
\eqnn\gaux
$$\eqalignno{
&A_{11}(z)=\nord(i\a_+\a_1\cdot\der\vp(z))^2\nord-\inv{3}\nord(i\der X(z))^2
\nord\,,\cr
\apar
&A_{12}(z)=\nord(i\a_+\a_1\cdot\der\vp(z))\cdot
(i\a_+\a_2\cdot\der\vp(z))\nord\,,\cr
\apar
&A_{22}(z)=\nord(i\a_+\a_2\cdot\der\vp(z))^2\nord\,,\cr
\apar
&B_1(z)=\ha i\a_+\a_1\cdot\der^2\vp(z)+\ha\nord(i\der X(z))^2\nord\,,\qquad
B_2(z)=\ha i\a_+\a_2\cdot\der^2\vp(z)\,,\cr
\apar
&C(z)=\sum_{\b\in\D(su(3))}
                \nord\exp[i\sqrt{2}\b\cdot X(z)]\nord\,.&\gaux\cr}
$$
The integrations over $y$ and $z_{1,p}$ $(p=1,2,3)$
can be done by use of the formulae in appendix  leading to
\eqn\gtypefin{\eqalign{
\hat T_s(z)&\propto \frac{g}{2}\sum_{i,j=1}^2 M_{ij}A_{ij}(z)
+\sum_{i=1}^2 m_iB_i(z)+C(z)\,.}}
This can readily be shown to be equivalent to \brstexact\
with $T^{\rm (pf)}(z)$ given by \gstensorii.

\subsec{Topological Conformal Field Theories --
Physical Observables and Correlation Functions}

We now turn to the problem of finding physical observables and their
correlation functions.
A prerequisite for some field to be a physical observable
would be the nullity of its conformal weight with respect to
$\hat T(z)$. We can easily spell out such fields among
$\Ps^{\La,\m}_\n(z)$'s:
\eqn\?{\mathop{\cup}\limits_{w\in W}\CN_w\,,\quad\hbox{where}
\quad \CN_w=\left\{\Ps_{w(\La+\r)-\r}^{\La,w(\La+\r)-\r}(z),\quad
\La\in P_+^k\right\}\,.}
Notice that these set of fields can be obtained from
the Ramond ground fields in sect.3 by simply shifting the exponential
factors of the bosons.  As a consequence of this, the
charge conjugation symmetry of the Ramond ground states translates
into the conjugation symmetry under the background charge $-2\a_0\r$,
{\it i.e.} $q\Leftrightarrow 2\a_0\r-q$.
More explicitly,
\eqn\?{\Ps_{w(\La+\r)-\r}^{\La,w(\La+\r)-\r}(z)
\mathop{\iff}\limits_{\rm conjugation}
\Ps_{ww_0(\bar\La+\r)-\r}^{\bar\La,ww_0(\bar\La+\r)-\r}(z)\,,}
and hence the conjugate set of $\CN_w$ is $\CN_{ww_0}$.
If we could adopt the set $\CN_w\cup\CN_{ww_0}$ for a fixed $w\in W$ as
physical observables of our topological field theory,
we would meet no difficulty in finding correlation
functions since due to the existence of the charge conjugation under the
background charge $-2\a_0\r$, we can for instance consider
\eqn\duality{\vev{\Ps_{w(\La+\r)-\r}^{\La,w(\La+\r)-\r}
 \Ps_{ww_0(\bar\La+\r)-\r}^{\bar\La,ww_0(\bar\La+\r)-\r}}_{-2\a_0\r}=1\,,}
However there exists a nuisance in this.
In general, two fields in $\CN_w\cup\CN_{ww_0}$ are not relatively
local and hence it will not be possible  to equip
$\CN_w\cup\CN_{ww_0}$
with a good ring structure.

We will consider in the following
\eqn\?{\CN_{w=\rm id}=\{\P_\La(z)=
\Ps^{\La,0}_\La(z)\ \big\vert\  \La\in P_+^k\}=\CR_{w=\rm id}\,,}
as physical observables. A nice thing about $\CN_{w=\rm id}$ is that
its elements $\P_\La$, $\La\in P_+^k$ form a ring  and obey
\eqn\physcond{\oint_0 dz G^{\a_i}(z)\P_\La(0)=0\,,\quad i=1,\ldots,n\,,}
as we have seen in sect.3. On the other hand this choice of physical
observables
apparently breaks conjugation symmetry and hence  we have to devise a
procedure to make viable correlation functions.
If we have a legitimate procedure to make correlation functions of $\P_\La$'s
the conditions \physcond\ and \brstexact\ imply, as in usual topological field
theories, that the correlation functions are independent of the world-sheet
positions of $\P_\La$'s.

Now we wish to tackle the problem of constructing correlation functions among
$\P_\La$'s to make our topological field theory
sensible. But before doing that, we should spend a while to think over what
the BRST structure of the theory must be. The problem of finding out the
 BRST-structure in our theory is essentially the same\foot{
However there is a slight  difference between the two models
as we shall explain shortly.}\
as that of extending the results of Felder \rF\ and others
\rOthers\ to the coset model
$G_k\times G_0/G_k$ using the representation in terms of free bosons and
parafermions. The latter problem is a formidable task and as far as we know
there seems to have been no complete analysis yet. Here we can only make
several
conjectural statements which we admit are not based on a rigorous footing.
Let $\CF_{\La,\La'}^\b$ be the space of the fields with  parafermionic
charge $\b$ and $u(1)^n$ charge $-\inv{\a_+}(\a_+\La+\a_-\La')$.
The `BRST charges' or the intertwiners
\eqn\?{Q_{12}:\CF_{\La_1,\La'_1}^{\b_1}
\longrightarrow\CF_{\La_2,\La'_2}^{\b_2}}
will be integrals of products of $G^{\a_i}(z)$'s but their explicit form
 must be dependent on $\CF_{\La_1,\La'_1}^{\b_1}$ and
$\CF_{\La_2,\La'_2}^{\b_2}$.
The spaces $\CF_{\La,\La'}^\b$ and their intertwiners determine a directed
graph
which we suppose coincides with the Hasse diagram of the affine Weyl group
of $\bg$. We conjecture that a part of this diagram is given by
\medskip
\eqn\diagram{
\matrix{
&\raise 5mm\hbox{$\mkern 100mu\searrow$}&
&\buildrel{\raise 2mm\hbox{$\mkern -20mu\scriptstyle\oint G^{\a_1}$}}
\over\nearrow
&\raise 9mm\hbox{$\CF^{\a_1}_{-\a_1,\La}$}\cr
\CF^{-\t}_{\t,\La}&\buildrel {\oint (G^{\a_1})^{a_1}\cdots (G^{\a_n})^{a_n}}
\over{\hbox to 3cm{\rightarrowfill}}&\CF^0_{0,\La}&\lower .5mm\hbox{$\vdots$}&
\cr
&\lower 6mm\hbox{$\mkern 100mu\nearrow$}&
&\lower 8mm\hbox{$\buildrel{\raise 2mm\hbox{$\displaystyle\searrow$}}
\over{\mkern -20mu\scriptstyle\oint G^{\a_n}} $}
&\lower 1.2cm\hbox{$\CF^{\a_n}_{-\a_n,\La}$}\cr}}
\bigskip
\noindent{where} we are again sloppy about the contours and the normalization
constants of the intertwiners.
For instance we have $\hat T(z)\in \CF_{0,0}^0$ and
$G^{-\t}(z)\in \CF^{-\t}_{\t,0}$.
If we could complete the diagram with the explicit knowledge of the
intertwiners,
we could construct the BRST complex and discuss its cohomology.
As mentioned above this is far beyond the scope of the present paper.
Our physical observables $\P_\La$'s are in  $\CF_{0,\La}^0$  and annihilated
by
$\oint G^{\a_i}$'s. We expect them to be non-trivial cocycles of the BRST
complex.

The difference between our topological field theory and the coset model
$G_k\times G_0/G_k$ resides in the following points. In the conventional
$G_k\times G_0/G_k$ model the generators of the chiral algebra are
Weyl-invariant
(like the stress-energy tensor or $\CW$ generators) and in particular
$u(1)^n$-charges are not considered as quantum numbers. Consequently
all the $\CN_w$'s with different $w$ are physically identified and the
`Kac table' is  labeled solely by $P_+^k$. It is believed, however, that in
practical calculations of correlation functions one may take any convenient
representatives  from appropriate $\CN_w$'s so as to minimize introductions
of the screening charges. On the other hand our topological field theory cares
about $u(1)^n$ quantum numbers since it was obtained by twisting the
higher-rank supersymmetric models which have $u(1)^n$ symmetry.
Therefore we have to consider correlation functions strictly within
$\CN_{\rm id}$ once we choose $\CN_{\rm id}$ as physical observables.

Since, by this restriction of physical observables, the original conjugation
symmetry is lost and it is inevitable to introduce
\eqn\?{\oint dzS^{\a_i}(z)\,,\quad \hbox{where
$S^{\a_i}(z)=\ps_{-\a_i}(z)\nord\exp[i\a_-\a_i\cdot\vp(z)]\nord$}\,,}
which however do not break the BRST invariance since
\eqn\?{\[\oint dz\,G^{\a_i}(z)\, , \,\oint dz\,S^{\a_j}(z)\]=0\,,}
by virtue of
\eqn\?{G^{\a_i}(z)S^{\a_i}(z')=\inv{(z-z')^2}\nord
e^{i\a_0\a_i\cdot\vp(z')}\nord
+\frac{\a_+}{\a_0}\inv{z-z'}\der_{z'}\(\nord e^{i\a_0\a_i\cdot\vp(z')}\nord\)
+\cdots\,,}
and can be used to balance the excess of the $u(1)^n$ charge caused by
products of  $\P_\La$'s. For instance,
\eqn\?{\vev{\(\oint S^{\a_1}\)^{m_1}\cdots\(\oint S^{\a_n}\)^{m_n}\P_\La
       \P_{\bar\La}}=1\,,}
where
\eqn\?{\La+\bar\La=\La-w_0(\La)=\sum_{i=1}^n m_i\a_i\,.}
In this way by inserting $\oint S^{\a_i}$'s one can consider an arbitrary
correlation functions of $\P_\La$'s
and hence one is able to formulate  a consistent topological field
theory (modulo some technical difficulties).

According to the general formalism of topological field theory
\refs{\rW,\rEY,\rDVV}, if we have BRST-invariant operators of  $(0,0)$-form
 there
exist  $(1,1)$-forms
(or (1,0)-forms if we restrict ourselves to the left chiral part) related via
`descent equations' whose integrals become again BRST-invariant.
Introduction of these extra physical observables can formally be interpreted
as
perturbations of the original topological field theory by the $(1,1)$-form
operators.
An analogous situation arises here for our topological field theory obtained
above.
Let
\eqn\?{\Ps_\La(z)=
\p^\La_{\La-\t}(z)\nord\exp[-i(\a_+\t+\a_-\La)\cdot\vp(z)]\nord}
for $\La\in P_+^k$. It is easy to see that $\Ps_\La(z)$
has a conformal weight 1 with respect to $\hat T(z)$ and $\Ps_\La(z)\in
\CF_{\t,\La}^{-\t}$.
Furthermore we have
\eqn\?{G^{-\t}(z)\P_\La(w)=\inv{z-w}\Ps_\La(w)+\cdots\,,}
and hence
\eqn\?{\eqalign{
&\oint (G^{\a_1})^{a_1}\cdots (G^{\a_n})^{a_n}\oint_0dw\Ps_\La(w)\cr
\apar
&=\oint (G^{\a_1})^{a_1}\cdots (G^{\a_n})^{a_n}\oint_0dw\oint_w dz
G^{-\t}(z)\P_\La(w)\cr
\apar
&\propto\oint_0dw\oint_w dz \hat T(z)\P_\La(w)=0\,,\cr}}
which shows that $\oint_0dw\Ps_\La(w)$, $\La\in P_+^k$ are BRST invariant
observables provided that the diagram \diagram\ is correct.

Here we should comment again on  the exceptional case $\bg=su(2)$.
 We noted in sect.3 that $\CV_{w=+1}$ is charge-conjugation symmetric per se,
{\it i.e.\/}
$\CV_{w=+1}=\CV_{w=-1}$. Similarly, we have here $\CN_{w=+1}=\CN_{w=-1}$.
In other words the chiral ring $\CN_{w=+1}=\CR_{w=+1}$ is symmetric under
conjugation with background charge $-2\a_0\r$ which is just
a manifestation of the Poincar{\'e}
duality in $N=2$ superconformal field theory. Therefore  it is
possible to consider correlation functions among the elements of $\CN_{w=+1}$
as in \duality\ without introducing an extra machinery. This is the route
usually taken in constructing  topological field theory from
the \minimal\ \rEY.   So we can obtain two possible topological field theories
starting from the \minimal, the usual one and the one constructed here.
They are different theories and this is particularly evident from the fact
that  their selection rules of physical observables  are
different.

As a final comment of this section we should  like to speculate on a possible
connection of our topological field theory to  $G/G$ topological field theory.
In the existing literature \rGoverG, $G/G$ theory is formulated by gauging
 the WZW model. Its physical observables are labeled by $P_+^k$ as in
our theory and  it has a marked property to the effect that
three-point functions essentially coincide with the structure constants of
the fusion algebra of the WZW model. We suspect that three-point functions
in our topological field theory
\eqn\threeptfunc{N_{\La\La'\La''}=\vev{\(\oint S^{\a_1}\)^{m_1}
\cdots\(\oint S^{\a_n}\)^{m_n}\P_\La\P_{\La'}\P_{\La''}}\,,}
where
\eqn\?{\La+\La'+\La''=\sum_{i=1}^n m_i\a_i\,,}
may similarly coincide with the fusion rules coefficients of the WZW
model and hence the two theories may be equivalent.
We remark that a similar idea is proposed  when the relation between the
twisted $N=2$ Kazama-Suzuki model and the $G/G$ theory is discussed \rNW.
Unfortunately the explicit evaluation of \threeptfunc\ seems quite a
formidable
task. In addition the two theories are formulated quite differently,
one in terms of the gauged WZW model and the other in terms of a combined
system
of parafermions and free bosons.
Thus whether this equivalence is indeed true or not remains open.

\newsec{Concluding Remarks}

In this paper we have elaborated on  the structure of higher-rank
supersymmetric
models which are natural generalization of the \minimal\ and have
endeavored to construct a class of topological field theories.
Although there remain technical problems to be overcome for our topological
field theories to become full-fledged, we hope we have convinced the reader
that such topological field theories can really exist.

We would like to end this paper with the following remark.
When the rank of $\bg$ exceeds one,
 the existence of
the extended Virasoro algebra is expected in our topological field theory.
Suppose that the extended algebra is  of the $\CW$ algebra type, then
the $\CW$ currents in topological algebra will take  the BRST-exact form.
Let us report  a tiny  calculation  which exemplifies this idea.
We start with a  BRST-exact expression
\eqn\?{
\CW(z)=\oint (G^{\a_1})^{a_1}\cdots(G^{\a_n})^{a_n}
\nord v\cdot J(z) G^{-\t}(z)\nord\,.}
It is easy to see that $\CW(z)$ is a spin-3 field with respect to
$\hat T(z)$ provided that
\eqn\Wcond{(2\a_0\r+\a_+\t)\cdot v=0\,.}
In the case $\bg=su(3)$, the condition \Wcond\ fixes $v$ uniquely up to
a multiplicative constant and via an explicit calculation\foot{
with the aid of {\it Mathematica}}\ 
we have confirmed for $k=1$
\eqnn\?
$$\eqalignno{
\CW(z)\propto &\nord\Biggl[ (\der\vp_2(z))^3-3\der\vp_2(z)(\der\vp_1(z))^2
+\frac{3i}{2\sqrt{2}}
\(\der\vp_1(z)\der^2\vp_1(z)-\der\vp_2(z)\der^2\vp_2(z)\)\cr
\apar
&\hskip 7mm
+\frac{\sqrt{6}i}{2}\der\vp_2(z)\der^2\vp_1(z)-\inv{8}\der^3\vp_2(z)
+\frac{\sqrt{3}}{8}\der^3\vp_1(z)\Biggr]\nord\,,&\?\cr}
$$
where the RHS is the standard spin-3 generator of the $\CW_3$ algebra
\refs{\rFZ,\rR}.
Thus in this case we obtain the $\CW_3$ extended topological algebra.
It is   a frequently asked but never completely answered question
precisely in what way the extended chiral algebra such as the $\CW$ algebra
and the idea of topological conformal field theory can be combined.
We hope our higher-rank topological models can  shed a new light on this issue
in the future.

\ack

One of us (SKY) would like to thank H.~Saleur for discussions and warm
hospitality at Yale University where  part of this work
was done.

\rappendix{A}{Selberg-type Integrals}

In this appendix we gather useful formulas for integrals of the form\foot{
{\it cf.} refs.~\refs{\rDF,\rA}.}
\eqn\Sel{\CJ^{(N)}(a_1,\ldots,a_N;b;c)=\oint\ts\prod_{i=1}^N dz_i
\prod_{i=1}^N z_i^{a_i-1} (1-z_i)^{b-1}\prod_{i<j}^N (z_i-z_j)^{2c}\,,}
where the integration contour must be closed and such that it  avoids the
singularities of the integrand.
It is easy to show that
\eqn\Zenm{\sum_{i=1}^N\[(a_i+m)\vev{z_i^m}-(b-1)
\vev{{\frac{z_i^{m+1}}{1-z_i}}}\]+
\sum_{i<j}^N 2c\vev{\frac{z_i^{m+1}-z_j^{m+1}}{ z_i-z_j}}=0,}
where we introduced the notation,
\eqn\Mom{\vev{f(z_1,\ldots,z_N)}=\oint\ts\prod_{i=1}^N dz_i
\prod_{i=1}^N z_i^{a_i-1} (1-z_i)^{b-1}\prod_{i<j}^N (z_i-z_j)^{2c}
f(z_1,\ldots,
z_N)\,,}
for any $f(z_1,\ldots,z_N)$.
By setting $m$=0  in \Zenm\  we find
\eqn\Mzero{(b-1)\sum_{i=1}^N\dddvev{{\inv{1-z_i}}}=
\sum_{i=1}^N a_i+N(b-1)+N(N-1)c\,}
where $\dvev{f(z_1,\ldots,z_N)}:=\vev{f(z_1,\ldots,z_N)}/\vev{1}$.
Using \Mzero,  eq.\Zenm\ can be rewritten as
\eqn\Zem{\eqalign{\sum_{i=1}^N&\[(a_i+m)\dvev{z_i^m}+(b-1)
\dddvev{{\frac{1-z_i^{m+1}}{1-z_i}}}\]\cr
\apar
&\hskip 1cm +\sum_{i<j}^N 2c\dddvev{\frac{z_i^{m+1}-z_j^{m+1}}{ z_i-z_j}}
=\sum_{i=1}^N a_i+N(b-1)+N(N-1)c\,.\cr}}
When $m$ is a positive integer greater than $1$, this becomes
\eqn\Zemp{\eqalign{\sum_{i=1}^N&\[a_i+b+2(N-1)c+m-1\]\dvev{z_i^m}
+(b-1)\sum_{i=1}^N \sum_{p=1}^{m-1}\dvev{z_i^p}\cr
\apar
&\hskip 1cm +2c\sum_{i<j}^N \sum_{p=1}^{m-1}\ddvev{z_i^p z_j^{m-p}}
=\sum_{i=1}^N a_i+N(N-1)c\,,\cr}}
while  in the case $m=1$ we find
\eqn\Zone{\sum_{i=1}^N(a_i+b+2(N-1)c)\dvev{z_i}=\sum_{i=1}^N a_i+N(N-1)c\,.}
For a fixed $j$ such that $1\le j \le N$ replace  $a_j$ by $a_j+1$ in \Zone.
Then we obtain
\eqn\Zod{\eqalign{\sum_{i\ne j}^N\(a_i+b+2(N-1)c\)\dvev{z_i z_j}
&+\(a_j+b+2(N-1)c+1\)\dvev{z_j^2}\cr
\apar
&=\[\sum_{i=1}^N a_i+N(N-1)c+1\]\dvev{z_j}.\cr}}
Eqs.\Zemp\ with $m$=2 and \Zod\
produce
\eqnn\Cod
$$\eqalignno{\sum_{i<j}^N&\[a_i+a_j+2b+2(2N-3)c\]\dvev{z_i z_j}\cr
\apar
&\hskip 15mm =\[\sum_{j=1}^N a_j+b+N(N-1)c\]\sum_{i=1}^N\dvev{z_i}
-\[\sum_{i=1}^N a_i+N(N-1)c\]\,.\cr
&&\Cod\cr}
$$
If we take  $a_i=a$ for all $i\ (i=1,\ldots,N)$, it then easily follows
that
\eqna\sexp
$$\eqalignno{&\sum_{i=1}^N\dvev{z_i}=\frac{N(a+(N-1)c)}{a+b+2(N-1)c}\,,
&\sexp a\cr
\apar
&\sum_{i<j}^N\dvev{z_i z_j}=\frac{\inv{2}N(N-1)(a+(N-1)c)(a+(N-2)c)}
{(a+b+(2N-2)c)(a+b+(2N-3)c)}\,,&\sexp b\cr
\apar
&\sum_{i=1}^N\dvev{z_i(1-z_i)}\cr
\apar
&\hskip 5mm =\frac{N(a+(N-1)c)(b+(N-1)c)(a+b+(N-2)c)}
{(a+b+2(N-1)c+1)(a+b+2(N-1)c)(a+b+(2N-3)c)}\,,\cr
\apar
&&\sexp c\cr}
$$
and
\eqnn\?
$$\eqalignno{\sum_{i<j}^N\dvev{(z_i-z_j)^2}&=(N-1)\sum_{i=1}^N\dvev{z_i^2}-2
\sum_{i<j}^N\dvev{z_i z_j}\cr
\apar
&=\frac{N(N-1)(Nc+1)(a+(N-1)c)(b+(N-1)c)}{(a+b+2(N-1)c+1)(a+b+2(N-1)c)
(a+b+(2N-3)c)}\,.\cr
\apar
&&\?\cr}
$$
For $N=1$,  $\CJ^{(N)}(a_1,\ldots,a_N;b;c)$ reduces to the beta function
$B(a,b)$ and eqs.\sexp{a}\
and  \sexp{c}\ become the well-known recursion formulas,
\eqna\?
$$\eqalignno{&B(a+1,b)=\frac{a}{a+b}B(a,b)\,,&\? a\cr
\apar
&B(a+1,b+1)=\frac{ab}{(a+b)(a+b+1)}B(a,b)\,.&\? b\cr}
$$

\listrefs
\vfil\eject

\inserttabtex{${\bf g}$\vt$g$\vt {\bf Dynkin diagrams} \eltt||\el
$\tphan\CA_n$|$n+1$|\raise -6mm\hbox{\ynepsf{adyn.ps}{6.6cm}}\el||\el
$\tphan\CB_n$|$2n-1$|\raise -6mm\hbox{\ynepsf{bdyn.ps}{6.6cm}}\el||\el
$\tphan\CD_n$|$2n-2$|\raise -7mm\hbox{\ynepsf{ddyn.ps}{6.6cm}}\el||\el
$\tphan\CG_2$|$4$|\raise -6mm\hbox{\ynepsf{g2dyn.ps}{1.43cm}}\el||}
{Relevant Dynkin diagrams and the numbering of their nodes.}
\vfil
\inserttabtex{
${\bf g}$\vt $a_i$ \eltt|\el
$\tphan\CA_n$|\raise -6mm\hbox{\ynepsf{akac.ps}{7.2cm}}\el|\el
$\tphan\CB_n$|\raise -6mm\hbox{\ynepsf{bkac.ps}{7.2cm}}\el|\el
$\tphan\CD_n$|\raise -7mm\hbox{\ynepsf{dkac.ps}{7.2cm}}\el|\el
$\tphan\CG_2$|\raise -6mm\hbox{\ynepsf{g2kac.ps}{1.56cm}}\el|}
{List of $a_i$'s where $\t=\sum_{i=1}^na_i\a_i$.}
\vfil

\midinsert
\vbox{
\begintable
${\bf g}$\vt$ M_{ij}\quad(1\le i\le j\le n)$\vt
$2\r=\sum_{i=1}^nm_i\a_i$ \eltt||\el
$\CA_n$
|$M_{ij}=\frac{i(n+1-j)}{n+1}$\hfill
|$2\r=\sum\limits_{i=1}^n i(n+1-i)\a_i$\hfill\el
||\elt
||\el
$\CB_n$
|$M_{ij}=i$\hfill
|$2\r=\sum\limits_{i=1}^ni(2n-i)\a_i$\hfill\el
||\elt
||\el
\hfill$\CD_n$
|\lower 9mm \vbox{
\+&$M_{ij}=i,\quad 1\le i,j\le n-2$\cr
\+&\vphantom{\vrule height 2mm}\cr
\+&$M_{i,n-1}=M_{i n}=\frac{i}{2},\quad 1\le i\le n-2$\cr
\+&\vphantom{\vrule height 2mm}\cr
\+&$M_{n-1,n-1}=M_{n n}=\frac{n}{4}$\cr
\+&\vphantom{\vrule height 2mm}\cr
\+&$M_{n-1,n}=\frac{n-2}{4}$\cr
}\hfill
|\lower 5mm \vbox{
\+&$2\r=\sum\limits_{i=1}^{n-2}2\(in-\frac{i(i+1)}{2}\)\a_i$\cr
\+&\vphantom{\vrule height 2mm}\cr
\+&\hskip 5mm$+\frac{n(n-1)}{2}(\a_{n-1}+\a_n)$\cr}\hfill\el
||\elt
||\el
$\CG_2$
|$M_{11}=6,\quad M_{12}=3,\quad M_{22}=2$\hfill|
$2\r=10\a_1+6\a_2$\hfill\el
||
\endtable}
\global\advance\tabno by1
\setbox0=\vbox{\noindent\leftskip=.2\hsize\rightskip=.2\hsize
{\bf Table \the\tabno}.  Some useful Lie algebraic data. }
\vskip 18pt
\centerline{\box0\hss}
\vskip 24pt
\endinsert
\vfil
\bye